\documentclass[11pt]{article}

\usepackage[subtle]{savetrees}

\usepackage[margin=1in]{geometry}
\usepackage{graphicx}
\usepackage{amsmath}
\usepackage{amssymb}
\usepackage{amsthm}
\usepackage{algorithmicx, algorithm}
\usepackage{algpseudocode}
\usepackage{mathtools,comment}
\usepackage{physics}
\usepackage{qcircuit}
\usepackage{xcolor,soul}
\setstcolor{red}
\usepackage{subcaption,xspace}
\usepackage{array}
\usepackage{thmtools} 
\usepackage{thm-restate}
\usepackage{accents}
\usepackage[affil-it]{authblk}

\usepackage{lineno}

\newtheorem{problem}{Problem}
\newtheorem{theorem}{Theorem}
\newtheorem{lemma}{Lemma}
\newtheorem{corollary}{Corollary}
\newtheorem{proposition}{Proposition}

\newcommand{\etalgo}{\texttt{HighDist-Algo}\xspace}
\newcommand{\highampalgo}{\texttt{HighAmp-Algo}\xspace}

\newcommand{\hdq}{\texttt{HD$_q$}\xspace}
\newcommand{\cmp}{\texttt{CMP}\xspace}
\newcommand{\pmax}{p_{max}\xspace}

\newcommand{\kd}{$k$-distinctness\xspace}

\newcommand{\D}{\mathcal{D}\xspace}
\newcommand{\Ot}{\Tilde{O}}

\newcommand{\prob}{\textsc{HighDist}\xspace}
\newcommand{\amprob}{\textsc{HighAmp}\xspace}

\newcommand{\kdist}{$k$-\textsc{Distinctness}\xspace}
\newcommand{\gkd}{\textsc{Gapped $k$-Distinctness}\xspace}
\newcommand{\dgkd}{$\Delta$-\gkd\xspace}
\newcommand{\Finf}{$\textsc{F}_\infty$\xspace}
\newcommand{\minent}{\textsc{Min-Entropy}\xspace}
\newcommand{\countdec}{\textsc{CountDecision}\xspace}
\newcommand{\ED}{\textsc{ElementDistinctness}}
\newcommand{\logn}{r}

\newcommand{\pmaxprob}{$\mathtt{P_{\max}}$\xspace}
\newcommand{\promiseprob}{\textit   {Promise}-\textsc{HighDist}\xspace}

\newcommand{\deltaerr}{\log\tfrac{1}{\delta\tau}}
\newcommand{\liwualgo}{{\tt LiWuAlgo}}

\newcommand{\eqae}{EQAmpEst\xspace}
\newcommand{\cuquad}{\qquad\qquad\qquad\qquad\quad}
\newcommand{\pmset}{S}
\newcommand{\fmax}{\hat{f}_{max}\xspace}

\newcommand{\simulaeprob}{\texttt{SimulAEProb}\xspace}
\newcommand{\simulalgo}{\texttt{SimulAE-Algo}\xspace}
\newcommand{\iden}{\mathbb{I}}
\newcommand{\mbc}[1]{\mathbf{C}^{(#1)}}
\newcommand{\tring}[1]{\accentset{\circ}{#1}}

\newcolumntype{C}[1]{>{\centering\let\newline\\\arraybackslash\hspace{0pt}}m{#1}}
\newtheorem*{lemmanonum}{Lemma}{\bfseries}{\itshape}

\algnewcommand\algorithmicon{\textbf{on}}
\algdef{SE}[ON]{On}{EndOn}[1]{\algorithmicon\ #1\ \algorithmicdo}{\algorithmicend\ \algorithmicloop}

\begin{document}
\title{\prob Framework: Algorithms and Applications}
%
%
\author{Debajyoti Bera ~~ Tharrmashastha SAPV}
\affil{Department of Computer Science, IIIT-D, India}


%
%
%
\date{\vspace{-5ex}}

{\nolinenumbers \maketitle}
\begin{abstract}
We introduce the problem of determining if the mode of the output distribution of a quantum circuit (given as a black-box) is larger than a given threshold, named \prob, and a similar problem based on the absolute values of the amplitudes, named \amprob. We design quantum algorithms for promised versions of these problems whose space complexities are logarithmic in the size of the domain of the distribution, but the query complexities are independent. 

Using these, we further design algorithms to estimate the largest probability and the largest amplitude among the output distribution of a quantum black-box. All of these allow us to improve the query complexity of a few recently studied problems, namely, $k$-distinctness and its gapped version, estimating the largest frequency in an array, estimating the min-entropy of a distribution, and the non-linearity of a Boolean function, in the $\Ot(1)$-qubits scenario. The time-complexities of almost all of our algorithms have a small overhead over their query complexities making them efficiently implementable on currently available quantum backends.

\end{abstract}

\section{Introduction}
\label{sec:intro}


A quantum circuit is always associated with a distribution, say $\D$, over the observation outcomes~\footnote{We assume measurement in the standard basis in this paper, however, it should not be difficult to extend our algorithms to accommodate measurements in another basis.} that can, in principle, encode complex information. Given a threshold $\tau$, and a blackbox to run the circuit, it may be useful to know if there is any outcome with probability at least $\tau$. We denote this problem \prob. We also introduce \amprob that determines if the absolute value of the amplitude of any outcome is above a given threshold; even though this problem appears equivalent to \prob, however, an annoying difference crawls in if we allow absolute or relative errors with respect to the threshold. The most interesting takeaway from this work are $\Ot(1)$-qubits algorithms for the above problems whose query complexities and time complexities are independent of the size of the domain of $\D$.

The framework offered by these problems supports interesting tasks. For example, a binary search over $\tau$ (tweaked to handle the above annoyance) can be a way to compute the largest probability among all the outcomes --- we call this the \pmaxprob problem. Similarly, non-linearity of a Boolean function can be computed by finding the largest amplitude of the output of the Deutsch-Jozsa quantum circuit~\cite{Bera2021QuantumEstimation}.

Going further, we observed surprising connections of \prob and \pmaxprob to few other problems that have been recently studied in the realm of quantum algorithms, {\it viz.}, \kdist\cite{Ambainis2007QuantumDistinctness,Belovs2012Learning-Graph-BasedK-Distinctness}, \gkd~\cite{Montanaro2016TheMoments}, \minent~\cite{Li2019QuantumEstimation}, and \Finf~\cite{Montanaro2016TheMoments,Bun2018ThePolynomials}. Using the above framework we designed query and time-efficient quantum algorithms for those problems that require very few qubits, often exponentially low compared to the existing algorithms. \prob, \amprob and \pmaxprob being fundamental questions about blackboxes that generate a probability distribution, we are hopeful that space-bounded quantum algorithms with low query complexities could be designed for more problems by reducing to them.

An interesting outcome of this work is a unified study of the problems given above, each of which have received separate attention. For example, Li et al.~\cite{Li2019QuantumEstimation} recently considered the min-entropy estimation problem of a multiset which is equivalent to computing its \Finf, a problem studied just a few years ago by Montanaro~\cite{Montanaro2016TheMoments} and Bun et al.~\cite{Bun2018ThePolynomials}. 
We illustrate the reductions in Figure~\ref{fig:reductions}. (See Appendix~\ref{appendix:reductions} for details.)

\begin{figure}
    \centering
    \includegraphics[width=0.6\textwidth]{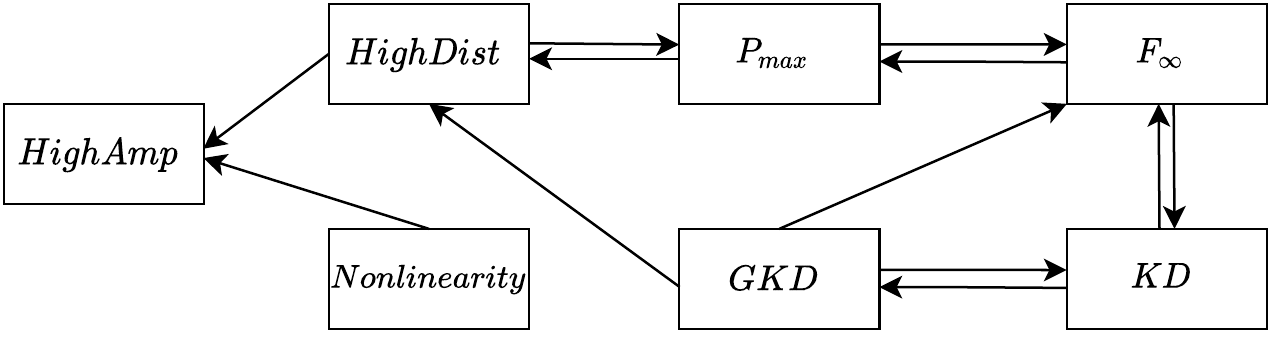}
    \caption{Reductions between the problems discussed in this paper.}
    \label{fig:reductions}
\end{figure}

The main contributions of this work can be summarized as follows.
\begin{enumerate}
    \item We introduce the \prob and the \amprob framework which allows us to answer interesting questions about the output distribution of a quantum circuit, like the largest probability, denoted \pmaxprob (a similar algorithm can also be designed for the largest absolute value among the amplitudes).
    
    \item We present space and query-efficient algorithms for the absolute and the relative error versions of the above problems. The algorithms for \prob and \pmaxprob are adapted from a recently published algorithm, and while they can be used to solve \amprob, we bettered their query complexities by designing a novel algorithm to run multiple amplitude estimations, in ``parallel'', and using a variant of the Hadamard test algorithm to estimate the inner-product of two output states.
    
    \item We show how to employ the above algorithms to improve the upper bounds on the query complexities of \kdist, \gkd, \minent, \Finf, and non-linearity estimation, all of which are now possible with logarithmic number of qubits --- often exponentially less compared to the existing approaches and leading to better space-time complexities. The reductions are mostly trivial, but the implications are interesting as discussed below.
    \begin{itemize}
        \item Our algorithm for \kdist makes optimal number of queries (up to logarithmic factors) when $k=\Omega(n)$, and that too using $\tilde{O}(1)$ qubits. Previous quantum algorithms for large $k$ have an exponential query complexity and require a larger number of qubits~\cite{Ambainis2007QuantumDistinctness}.
        \item Our algorithm for \prob can be used to identify the presence of high-frequency items in an array (above a given threshold --- also known as ``heavy hitters'') using $\tilde{O}(\log\tfrac{1}{\epsilon})$ qubits; it also generates a superposition of such items along with estimates of their frequencies. The best algorithms for identifying heavy hitters in low space classical algorithms are of streaming nature but require $\tilde{O}(\tfrac{1}{\epsilon})$ space~\cite{CMSketch}. Here $\epsilon\in(0, 1]$ indicates the inaccuracy in frequency estimation.
        \item Watson established the classical intractability of estimating min-entropy of a probabilistic source~\cite{watson2016complexity}; we show that the problem becomes easier for quantum algorithms when allowed to err in a small number of cases.
        \item Valiant and Valiant showed that $\tilde{O}(\tfrac{m}{\epsilon^2})$ samples of an $m$-valued array are sufficient to classically estimate common statistical properties of the distribution of values in the array~\cite{ValiantValiant}. Recently it was shown that fewer samples of the order of $\tilde{O}(\tfrac{1}{g^2})$ can be used if we want to identify the item with the largest probability (denoted $p_{\max}$)~\cite{Dutta2010ModeDistributions}; here $g$ denotes the gap between $p_{\max}$ and the second largest probability and is always less than $\pmax$. Our \pmaxprob quantum algorithm makes only $\tilde{O}(\tfrac{1}{g \sqrt{p_{\max}}})$ queries, finds the item and estimates its frequency with additive error.
        \item We recently showed that \prob and \pmaxprob can estimate non-linearity of any Boolean function with additive accuracy $\lambda$ using $\Ot(1)$ qubits and $\Ot(\tfrac{1}{\lambda^2 \fmax})$ queries~\cite{Bera2021QuantumEstimation}; here $\fmax$ denotes the largest absolute value of any Walsh coefficient of the function. Now, we can use $\amprob$ instead of \prob to do the same but using only $\Ot(\tfrac{1}{\lambda\fmax})$ queries. It should be noted in this context that the best known lower bound for non-linearity estimation is $\Omega(\tfrac{1}{\lambda})$~\cite{Bera2021QuantumEstimation}.
    \end{itemize}
\end{enumerate}

\begin{table}[!h]
\small
\centering
\caption{Results for the \kdist problem\label{table:summary-kdist}}
\begin{tabular}{ |C{2.5cm}||C{5.8cm}|C{6cm}|  }
 \hline
 \multicolumn{3}{|c|}{\kdist} \\
 \hline
 & Prior upper bound~\cite{Ambainis2007QuantumDistinctness}  & Our upper bound\\
 \hline
 $k\in\{2,3,4\}$ 
    & Setting $r = k$, $O((\frac{n}{k})^{k/2})$ queries, $O(\log(m)+\log(n))$ space &  $\tilde{O}(n^{3/2}/\sqrt{k})$ queries, $O\big((\log(m)+\log(n))\log(\frac{n}{\delta k})\big)$ space\\
\hline
$k=\omega(1)$ and $k\ge 4$ & $O(\frac{n^2}{k})$ queries,\newline $O(\log(m)+\log(n))$ space for $r\ge k$ &  $\tilde{O}(n^{3/2}/\sqrt{k})$ queries, $O\big((\log(m)+\log(n))\log(\frac{n}{\delta k})\big)$ space\\
 \hline
$k=\Omega(n)$ & $O(n^{n/2})$ queries,\newline $O(n\log(m) + \log(n))$ space &  $\tilde{O}(n)$ queries, $O\big((\log(m)+\log(n))\log(\frac{n}{\delta k})\big)$ space\\
\hline
\hline
\dgkd  & None & $\Tilde{O}(\frac{n^{3/2}}{\Delta\sqrt{k}})=\Tilde{O}((\frac{n}{\Delta})^{3/2})$ queries, $\Tilde{O}(1)$ space ($\Delta$ denotes additive error)\\
\hline
\end{tabular}
\end{table}


\begin{table}[!h]
\small
    \centering
    \caption{Algorithms for \Finf ($\epsilon < n$ denotes additive error)}
    \label{table:summary-finf-ub}
    \begin{tabular}{|p{6cm}|p{5.5cm}|p{3.5cm}|}
    \hline
    Approach & Query and\newline space complexity & Nature of error\\
    \hline
    binary search with $k$-distinctness~\cite[Sec~2.3]{Montanaro2016TheMoments} & $O(n \log(n))$ queries,\newline  $O(n)$ space & exact \\
    \hline
    $k$-distinctness with $k=\lceil \tfrac{16 \log(n)}{\epsilon^2} \rceil$~\cite{Li2019QuantumEstimation} & $O(n)$ queries,\newline  $O(n)$ space & $\epsilon$ additive error\\
    \hline
    quantum maximum finding over frequency table~\cite[Sec~3.3]{Montanaro2016TheMoments} & $O(n^{3/2})$ queries,\newline  $O(\log(m) + \log^2(n))$ space & exact\\
    \hline
    reducing to \pmaxprob \newline (binary search with \prob) [this] & $\Tilde{O}((n/\epsilon)^{3/2}\log(n/\epsilon))$ queries,\newline  $O\big((\log(m) + \log(\frac{n}{\epsilon}))\log(\frac{n}{\delta\epsilon})\big)$ space & $\epsilon$ additive error,\newline set $\epsilon=0.99$ for exact\\
    \hline
    \end{tabular}
\end{table}

\begin{table}[!h]
\small
    \centering
    \caption{Algorithms for non-linearity estimation ($\lambda$ denotes additive error)}
    \label{table:summary-nonlin-ub}
    \begin{tabular}{|p{3.2cm}|p{5.5cm}|p{6cm}|}
    \hline
    Approach & Query complexity & Space complexity\\
    \hline
    & &\\ [-1em]
    Using \prob~\cite{Bera2021QuantumEstimation} & $O(\frac{1}{\lambda^2\fmax}\log(\frac{1}{\lambda})\log(\frac{1}{\lambda\delta}))$ queries~$^\dagger$ &  $O\big((\log(n)+\log(\frac{1}{\lambda}))\cdot \log(\frac{1}{\delta\lambda})\big)$ space\\
    & & \\ [-1em]
    \hline
    & & \\ [-1em]
    Using \amprob & $O(\frac{1}{\lambda\fmax}\log(\frac{1}{\lambda})\log(\frac{1}{\lambda\delta}))$ queries & $O\big((\log(n)+\log(\frac{1}{\lambda}))\cdot \log(\frac{1}{\delta\lambda})\big)$ space\\
    \hline
    \multicolumn{3}{p{\linewidth}}{\footnotesize  $^\dagger$ Although the query complexity of the algorithm is presented as $\tilde{O}(\frac{1}{\lambda^3})$ queries in~\cite{Bera2021QuantumEstimation}, since we are merely estimating the largest probability in the output of the Deutsch-Jozsa algorith, using \pmaxprob gives us this tighter bound.}
    \end{tabular}
\end{table}


A summary of our results is presented in Tables~\ref{table:summary-kdist}, 
\ref{table:summary-finf-ub}, and \ref{table:summary-nonlin-ub}.  
Our algorithms work in the bounded-error setting and we shall often hide the $\log()$ factors in the complexities under $\tilde{O}()$. The time-complexities, except for \amprob, are same as the query complexities with logarithmic overheads since the techniques rely on quantum amplitude estimating, amplification and simple classical steps.

When space is not a constraint, query complexity of a problem for an $n$-sized array is $O(n)$ which is achievable by querying and caching the entire input at the beginning. However, this is not feasible when space is limited. This is also the scenario in the streaming setting, however, the focus there is to reduce the number of passes over the input under restricted space. In contrast, our algorithms are allowed only constant many logarithmic-sized registers, and they try to optimize the number of queries. To restrict the number of qubits to $\tilde{O}(1)$ we end up using super-linear queries for most of the problems. 
A rigorous space-time analysis can settle the tightness of those query complexities; we leave this direction open.

\section{Algorithms for \prob and \amprob}
Promise versions of the \prob problem play a central role in this work. 

\begin{restatable}[\prob]{problem}{hdprob}
\label{lemma:hdprob}
We are given a $(\log(m)+a)$-qubit quantum oracle $O_D$ that generates a distribution $D:\big(p_x = |\alpha_x|^2\big)_{x=1}^m$ upon measurement of the first $\log(m)$ qubits of \\\centerline{$\displaystyle O_D \ket{0^{\log(m)+a}} = \sum_{x \in \{0,1\}^{\log(m)}} \alpha_x \ket{x} \ket{\psi_x} = \ket{\Psi}$ (say)}\\ in the standard basis. We are also given a threshold $\tau \in (0,1)$ and the task is to identify any $x$ such that $p_x = |\alpha_x|^2 \ge \tau$, or report its absence. In the promise version with additive accuracy, we are given an additional $\epsilon \in (0,\tau)$, and the goal is to decide whether there exists any $x$ such that $p_x \ge \tau$ or if $p_x < \tau-\epsilon$ for all $x$, under the promise that only one of the cases is true. In the promise version with relative accuracy, the goal is to similarly decide between $p_x < (1-\epsilon_r)\tau$ and $p_x \ge \tau$, given some $\epsilon_r \in (0,1)$.
\end{restatable}

The algorithm for \prob follows these high-level steps.
\begin{itemize}
    \item Estimate $p_x=|\alpha_x|^2$ for all $x$ in another register, allowing relative or additive error as required, by employing vanilla quantum amplitude estimation (except the final measurement step, denoted QAE). This requires two copies of $\ket{\psi}$, one on which to operate the QFT-based circuit, and another, to furnish the ``good'' states (whose probabilities should be estimated).
    \item Compare each estimate $\ket{\Tilde{p_x}}$ with the threshold, hardcoded as $\ket{\tau}$. The comparison actually happens with a scaled version of $\tau$ since QAE does not generate $p_x$ directly. The states $\ket{x}$ for which $p_x \ge \tau$ are marked (in another register).
    \item The probability of finding a marked state, given there is one, is amplified using amplitude amplification. Care has to be taken to ensure that any $x$ for which $p_x < \tau-\epsilon$ but whose estimate is above $\tau$ is not sufficiently amplified.
\end{itemize}

The novelty of this workflow is the execution of QAE in parallel and a complex analysis showing that errors are not overwhelming. This is essentially the strategy followed by the {\tt QBoundFMax} quantum circuit that was recently proposed by us for estimating non-linearity~\cite[Algorithm~3]{Bera2021QuantumEstimation}.
We observed that {\tt QBoundFMax} can be repurposed based on the three following observations. First, {\tt QBoundFMax} identified whether there exists any basis state whose probability, upon observing the output of a Deutsch-Jozsa circuit, is larger than a threshold in a promised setting; however, no specific property of Deutsch-Jozsa circuit was being used. Secondly, amplitude estimation can be used to estimate $|\alpha_x|^2$ (with bounded error) in $\sum_x \alpha_x \ket{x} \ket{\xi_x}$ for any $x \in [n]$ by designing a sub-circuit on only the first $\log(m)$ qubits to identify ``good'' states (this sub-circuit was referred to as $EQ$ in {\tt QBoundFMax}).
Lastly, amplifying some states in a superposition retains their relative probabilities. These observations not only allow us to modify the {\tt QBoundFMax} algorithm for \prob, but also enable us to identify some $x$ such that $|\alpha_x|^2 \ge \tau$, along with an estimate of $|\alpha_x|^2$.

\begin{restatable}[Additive-error algorithm for \prob]{lemma}{etalgolemma}
\label{lemma:etalgo_correct}
Given an oracle $O_D$ for the \prob problem,
$m$ --- the domain-length of the distribution it generates, and a threshold $\tau$, along with parameters $0 < \epsilon < \tau$ for additive accuracy and $\delta$ for error, \etalgo is quantum algorithm that uses $O\big((\log(n)+\log(\frac{1}{\epsilon})+a)\log(\frac{1}{\delta\tau})\big)$ qubits and makes $O(\frac{1}{\epsilon\sqrt{\tau}}\deltaerr)$ queries to $O_D$. When its final state is measured in the standard basis, we observe the following. 
\begin{enumerate}
    \item If $p_x < \tau - \epsilon$ for all $x$ then the output register is observed in the state $\ket{0}$ with probability at least $1-\delta$.
    \item If $p_x \ge \tau$ for any $x$, then with probability at least $1-\delta$ the output register is observed in the state $\ket{1}$.
\end{enumerate}
\end{restatable}

It is reasonable to require that $\epsilon \ll \tau/2$, and in that case the query complexity can be bounded by $\tilde{O}(\tfrac{1}{\epsilon^{3/2}})$. The above algorithm can be converted to work with a relative accuracy by setting  $\epsilon = \epsilon_r \tau$.

\begin{restatable}[Relative-error algorithm for \prob]{lemma}{etalgolemmarel}
\label{lemma:etalgo_rel_correct}
There exists an algorithm to solve the promise version of \prob with relative inaccuracy $\epsilon_r$ in the similar manner as stated in Lemma~\ref{lemma:etalgo_correct} that makes $O(\frac{1}{\epsilon_r\tau^{3/2}}\deltaerr)$ queries to $O_D$ and uses $O\big((\log(m)+\log(\frac{1}{\epsilon_r\tau})+a)\log(\frac{1}{\delta\tau})\big)$ qubits.
\end{restatable}



\subsection{Algorithm for \amprob}
In the \amprob problem, the setup is same as that of \prob, but we are now interested to identify any $x$ such that $|\alpha_x| \ge \tau$. Though this is identical to \prob with threshold $\tau^2$, we have to set the threshold to $\tau^2$ and additive accuracy to $\epsilon^2$ if we want to use  Lemma~\ref{lemma:etalgo_correct} directly; this leads to query complexity $\Ot(\tfrac{1}{\epsilon^2 \tau})$. We design a new algorithm to improve upon this based on the observation that, despite the name, QAE actually estimates the {\em probability} of a ``good'' state; thus, why not estimate the amplitudes directly?
\begin{itemize}
    \item For all $x$ (in superposition), generate a state in another register which is $\ket{0}$ with probability $|\alpha_x|$. For this we designed an algorithm to essentially estimate the inner product of two states using a generalization of the Hadamard test, instead of the swap test.
    \item Employ amplitude amplification to estimate the probability of the state being $\ket{0}$, allowing relative or additive error as required. 
    To do this in superposition, i.e., for all $x$, with a low query complexity required us to design an algorithm for {\em simultaneous amplitude estimation}. The estimate is stored in another register as $\ket{\tilde{|\alpha_x|}}$.
    \item Compare each estimate $\ket{\Tilde{|\alpha_x|}}$ with the threshold, hardcoded as $\ket{\tau}$, and followup with similar steps as before.
\end{itemize}

\subsubsection{Hadamard test to estimate inner product of two states}
Say, we have two algorithms $A_{\psi}$ and $A_{\phi}$ that generate the states $A_{\psi}\ket{0^n}=\ket{\psi}$ and $A_{\phi}\ket{0^n}=\ket{\phi}$, respectively, and we want to produce a state $\ket{0}\ket{\xi_0} + \ket{1}\ket{\xi_1}$ such that the probability of observing the first register to be in the state $\ket{0}$ is linearly related to $|\braket{\psi}{\phi}|$. Though swap-test is commonly used towards this purpose, there the probability is proportional to $|\braket{\psi}{\phi}|^2$; this subtle difference becomes a bottleneck if we are trying to use amplitude estimation to estimate that probability with additive accuracy, say $\epsilon$. We show that the Hadamard test can do the estimation using $O(1/\epsilon)$ queries to the algorithms whereas it would be $O(1/\epsilon^2)$ if we use the swap test.

The Hadamard test circuit requires one additional qubit, initialized as $\ket{0}$ on which the $H$-gate is first applied. Then, we apply a conditional gate controlled by the above qubit that applies $A_{\psi}$ to the second register, initialized to $\ket{0^n}$, if the first register is in the state $\ket{0}$, and applies $A_{\phi}$ if the first register is in the state $\ket{1}$. Finally, the $H$-gate is again applied on the first register.

It is easy to calculate that the probability of measuring the first register as $\ket{0}$ is
$Pr\big[\ket{0}_{R_1}\big] = \Big\| \frac{1}{2} \big(\ket{\psi} + \ket{\phi}\big) \Big\|^2 = \frac{1}{2}\big(1-\big| \bra{\psi}\ket{\phi} \big|\big)$.
Thus, to obtain $|\bra{\psi}\ket{\phi}|$ with $\epsilon$ accuracy, it suffices to estimate $\frac{1}{2}\big(1-|\bra{\psi}\ket{\phi}|\big)$ with $\epsilon/2$ accuracy which can be performed by QAE using $O(1/\epsilon)$ queries to $A_{\phi}$ and $A_{\psi}$.

\subsubsection{Simultaneous Amplitude Estimation}
Let $[N] = \{y : 0\le y < 2^n-1 = N-1\}$ be an index set for some $n\in \mathbb{N}$.  
Suppose that we are given a family of quantum algorithms $\{A_y~:~y\in[N]\}$ each making $k$ queries to an oracle $O$, for some known constant $k$. Then for each $y$, $A_y$ can be expressed as $A_y = U_{(k,y)} O U_{(k-1,y)} \cdots U_{(1,y)} O U_{(0,y)}$ with suitable $U_{(i,y)}$ unitaries.
Let the action of $A_y$ on $\ket{0}$ be defined as $A_y\ket{0} = \beta_{0y} \ket{0} + \beta_{1y}\ket{1}$ denoted $\ket{\xi_y}$. (This can also be easily generalized if $A_y$s are $n$ qubit algorithms.)
Given an algorithm $A_{initial}$ to prepare the state $\ket{\Psi} = \sum_y\alpha_y\ket{y}$, the objective is to simultaneously estimate the ``probability'' of $\ket{0}$ in {\em each} $A_y\ket{0}$, i.e., obtain a state of the form $$\ket{\Phi} = \sum_y \alpha_y\ket{y}\ket{\xi_y}\ket{\tilde{\beta}_{0y}},$$ where, for each $y$, $\sin^2\Big(\frac{\tilde{\beta}_{0y} \pi}{2^m}\Big) = \Breve{\beta}_{0y}$ is an estimate of $\beta_{0y}$ such that $|\Breve{\beta_{0y}}-\beta_{0y}|\le \epsilon$ for some given $0<\epsilon\le 1$.


A naive approach to solve this problem would be to perform amplitude estimation of $\ket{0}$ in the state $\ket{\xi_y}$, conditioned on the first register being in $\ket{y}$, serially for each individual $y$.
Then, the total number of queries to the oracle $O$ would be $O(\frac{Nk}{\epsilon})$ where $O(k/\epsilon)$ is the query complexity due to a single amplitude estimation. However, we present an algorithm that performs the same task but with just $O(\frac{k}{\epsilon})$ queries to the oracle $O$. For this we require a controlled-version of the $\{A_y\}$ circuits. Let $A$ be an algorithm defined as $A = \sum_y \ketbra{y} \otimes A_y$ that operates $A_y$ on the second register if the first register is in $\ket{y}$.

We denote the amplitude estimation operator due to Brassard et al.~\cite{brassard2002quantum} as $AmpEst$. The operator to obtain an estimate with $m$ bits of precision can be expressed as $AmpEst = (F_m^{-1} \otimes \iden)\cdot \Lambda_m(G) \cdot (F_m \otimes \iden)$ where $F_m$ is the Fourier transform on $m$ qubits, $\Lambda_m(G)$ is the conditional operator defined as $\sum_x \ketbra{x}\otimes G^x$, $G = -AS_{\overline{0}}AS_{\chi}$ is the Grover operator and $G^x$ implies that the $G$ operator is applied $x$ times in succession.
Also let $AmpEst_y$ be defined as $AmpEst = (F_m^{-1} \otimes \iden)\cdot \Lambda_m(G_y) \cdot (F_m \otimes \iden)$ where $G_y = -A_yS_0A_y^{\dagger}S_\chi$. Then notice that $\ket{\Phi}$ can be obtained from $\ket{\Psi}$, as $$\ket{\Phi} = \Big(\sum_y \ketbra{y}\otimes AmpEst_y\Big)\Big(A\otimes I^{m}\Big)\cdot \ket{\Psi}\ket{0}\ket{0^m}.$$
By $\mathbf{U}$ we denote the operator $\sum_y \ketbra{y}\otimes AmpEst_y$. We show that $\mathbf{U}$ can be implemented using $O(k\cdot 2^m) = O(k/\epsilon)$ queries to the oracle $O$ at the expense of additional non-query gates which can even be exponential in $n$.

\begin{algorithm}[!h]
    \small
	\caption{\label{algo:simul_ae_algo}Simultaneous Amplitude Estimation Algorithm \simulalgo}
	\begin{algorithmic}[1]
	    \Require Oracle $O$, the set of indexed algorithms $\{A_y\}$, the algorithm $A_{initial}$, accuracy $\epsilon$ and error $\delta$.
	    \State Set $m = \lceil\frac{1}{\epsilon}\rceil+3$.
	    \State Initialize three registers $R_1R_2R_3$ as $\ket{0^n}\ket{0}\ket{0^m}$.
	    \State Apply $A_{initial}$ on $R_1$.
	    \State Apply $A=\sum_y\ketbra{y}\otimes A_y$ on $R_1R_2$.
	    \State Apply the quantum Fourier transform (QFT) $F_m$ on $R_3$.
	    \For{$i$ in $1$ to $m$, conditioned on $i^{th}$ qubit of $R_3$ being in $\ket{1}$, for $2^i$ many times}
	        \State Apply $S_{\chi}$
        	    \For{$j$ in $1$ to $k-1$}
            	    \For{$y$ in $0$ to $N-1$}
            	        \State Apply $U_{(j,y)}$ on $R_2$ conditioned on $R_1$ being $\ket{y}$.
            	    \EndFor
            	    \State Apply $O$ on $R_2$.
            	\EndFor
            	\For{$y$ in $0$ to $N-1$}
            	        \State Apply $U_{(k,y)}$ on $R_2$ conditioned on $R_1$ being $\ket{y}$.
        	        \EndFor
        	 \State Apply $S_{\overline{0}}$.
        	 \State Apply the transpose of the operations from line 8 to line 16 in reverse.
        \EndFor
        \State Apply the inverse QFT $F_m^{-1}$ on $R_3$.
        \State \Return $R_1R_2R_3$.
	\end{algorithmic}
    \end{algorithm}


\begin{restatable}[Simultaneous Amplitude Estimation]{theorem}{simulae_thm}
\label{theorem:simulaetheorem}
    Given an oracle $O$, a description of an algorithm $A = \sum_y\ket{y}\bra{y}\otimes A_y$ as defined earlier, an initial algorithm $A_{initial}$, an accuracy parameter $\epsilon$ and an error parameter $\delta$, \simulalgo uses $O(\frac{k}{\epsilon})$ queries to the oracle $O$ and with probability at least $1-\delta$ outputs
    $$\ket{\Phi} = \sum_y \alpha_y\ket{y}\ket{\xi_y}\ket{\tilde{\beta}_{0y}},$$ where $\sin^2\big(\frac{\tilde{\beta}_{0y}\pi}{2^m}\big)=\Breve{\beta_{0y}}$ is an $\epsilon$-estimate of $\beta_{0y}$ for each $y$.
\end{restatable}

Details of the \prob and \amprob algorithms and their analysis can be found in Appendix~\ref{appendix:hd_algo_1}, and those for the Hadamard test and simultaneous amplitude estimation can be found in Appendix~\ref{appendix:sae}.

\section{\pmaxprob and \minent problem}

\noindent The \pmaxprob problem is a natural extension of \prob.
\begin{problem}[\pmaxprob] Compute $\pmax = \max_{i\in [n]} p_i$ given a distribution oracle as required for the \prob problem.
\end{problem}

The min-entropy of a distribution $D=(p_i)_{i=1}^m$ is defined as $\max_{i\in[m]} \log(1/p_i)$ and the \minent problem is to estimate this value; clearly, estimating it with an additive accuracy is equivalent to estimating $\max_{i=1}^m p_i$ with relative accuracy. 
%
The currently known approach for this problem, in an array setting, involves reducing it to \kdist~\cite{Li2019QuantumEstimation} with a very large $k$, however, we show that we can perform better if we binary search for the largest threshold successfully found by the \prob problem.

\begin{restatable}[Approximating $\pmax$ with additive error]{lemma}{pmaxadditive}
\label{lemma:pmax_lemma_additive}
Given an oracle as required for the \prob problem, additive accuracy $\epsilon \in (0,1)$ and error $\delta$, there is a quantum algorithm that makes $O(\frac{1}{\epsilon\sqrt{\pmax}}\log{(\tfrac{1}{\epsilon})}\log(\frac{1}{\delta\cdot \pmax}))$ queries to the oracle and outputs an estimate $\widehat{\pmax}$ such that 
$|\pmax - \widehat{\pmax}| \le \epsilon$ with probability $1-\delta$. The algorithm uses $O\big((\log(m)+log(\frac{1}{\epsilon})+a)\log(\frac{1}{\delta\cdot\pmax})\big)$ qubits.

There is a similar algorithm that estimates $\pmax$ as $(1-\epsilon)^2 \widehat{\pmax} \le \pmax \le \widehat{\pmax}$ using $\tilde{O}(\frac{m^{3/2}}{\epsilon})$ queries on $O\big((\log(\frac{m}{\epsilon})+a)\log(\frac{1}{\delta\pmax})\big)$ qubits.
\end{restatable}

The algorithm for additive accuracy is essentially the {\tt IntervalSearch} algorithm that we recently proposed~\cite{Bera2021QuantumEstimation}. We further modified the binary search boundaries to adapt it for relative accuracy.




We are not aware of significant attempts to estimate $\pmax$ (or min-entropy) using a blackbox generating some distribution, except a result by Valiant and Valiant in which they showed how to approximate the distribution by a histogram~\cite{ValiantValiant} that requires $\tilde{O}(\tfrac{m}{\epsilon^2 \log m})$ samples, and another by Dutta et al~\cite{Dutta2010ModeDistributions} for finding the mode of an array. In the latter work the authors show that the modal element of $\tilde{O}(\tfrac{1}{g^2})$ samples from D is the modal element of $D$ with high probability, in which $g$ is the difference of the mode to the second highest frequency. Suppose we are given $g$ or some upper bound. Setting $\epsilon=\tfrac{g}{2}$ in Lemma~\ref{lemma:pmax_lemma_additive} allows us to obtain the modal element using $\tilde{O}(\tfrac{1}{g^{3/2}})$ queries. The former technique requires keeping $\tilde{O}(\tfrac{m}{\epsilon^2 \log m})$ elements, and the latter technique requires storage of $\tilde{O}(\tfrac{1}{g^{3/2}})$ elements (each element requires an additional $\log(m)$ bits); our technique, on the other hand, requires $O(\log(\tfrac{m}{\epsilon})\log(\tfrac{1}{\delta\pmax}))=O(\log(\tfrac{m}{g})\log(\tfrac{1}{\delta g}))$ qubits.



Details of the \pmaxprob algorithms and their analyses can be found in Appendix~\ref{appendix:pmax}.
 
\section{Problems based on arrays and Boolean functions}
 The algorithms for \kd, \dgkd, \Finf, and non-linearity estimation are obtained by reducing them to \prob or \pmaxprob (see Appendices \ref{sec:gkd}, \ref{sec:min_entropy_est}, and \ref{sec:non-lin-est} for details). A subtlety in those reductions is an implementation of $O_D$ given an oracle to an array --- this is explained in Appendix~\ref{subsec:oracle}.

\subsection{The \kdist and the \gkd problems}

The \textsc{ElementDistinctness} problem~\cite{Buhrman2005QuantumDistinctness,Ambainis2007QuantumDistinctness,Aaronson2004QuantumProblems} is being studied for a long time both in the classical and the quantum domain. It is a special case of the \kdist problem~\cite{Ambainis2007QuantumDistinctness,Belovs2012Learning-Graph-BasedK-Distinctness} with $k=2$. 

\begin{problem}[\kdist]
Given an oracle to an $n$-sized $m$-valued array $A$, decide if $A$ has $k$ distinct indices with identical values.
\end{problem}

By an $m$-valued array we mean an array whose entries are from $\{0, \ldots, m-1\}$. 
Observe that, \kdist can be reduced to \prob with $\tau=\tfrac{k}{n}$, assuming the ability to uniformly sample from $A$.

The best known classical algorithm for \kdist uses sorting and has a time complexity of $O(n\log(n))$ with a space complexity $O(n)$.
%
In the quantum domain, apart from $k=2$, the $k=3$ setting has also been studied earlier~\cite{Belovs2014ApplicationsAlgorithms,Childs2013AUpdates}.
The focus of all these algorithms has been primarily to reduce their query complexities. As a result their space requirement is significant (polynomial in the size of the list), and beyond the scope of the currently available quantum backends with a small number of qubits.
Recently Li et al.~\cite{Li2019QuantumEstimation} reduced the \minent problem to \kdist with a very large $k$ making it all the more difficult to implement.

The \kdist problem was further generalized to $\Delta$-\gkd  by Montanaro~\cite{Montanaro2016TheMoments} which comes with a promise that either some value appears at least $k$ times or every value appears at most $k-\Delta$ times for a given gap $\Delta$.  The \Finf problem~\cite{Montanaro2016TheMoments,Bun2018ThePolynomials} wants to determine, or approximate, the number of times the most frequent element appears in an array, also known as the modal frequency. Montanaro related this problem to the \gkd problem but did not provide any specific algorithm and left open its query complexity~\cite{Montanaro2016TheMoments}.
So it appears that an efficient algorithm for $\Delta$-\gkd can positively affect the query complexities of all the above problems. However, $\Delta$-\gkd has not been studied elsewhere to the best of our knowledge.

\subsection{Upper bounds for the \kdist problem}
The $k=2$ version is the \ED problem which was first solved by Buhrman et al.~\cite{Buhrman2005QuantumDistinctness}; their algorithm makes $O(n^{3/4}\log(n))$ queries (with roughly the same time complexity), but requires the entire array to be stored using qubits. A better algorithm was later proposed by Ambainis~\cite{Ambainis2007QuantumDistinctness} using a quantum walk on a Johnson graph whose nodes represent $r$-sized subsets of $[n]$, for some suitable parameter $r \ge k$. He used the same technique to design an algorithm for \kdist as well that uses $\tilde{O}(r)$ qubits and $O(r+(n/r)^{k/2}\sqrt{r})$ queries (with roughly the same time complexity). Later Belovs designed a learning-graph for the \kdist problem, but only for constant $k$, and obtained a tighter bound of $O(n^{\frac{3}{4}-\frac{1}{2^{k+2}-4}})$. It is not clear whether the bound holds for non-constant $k$, and it is often tricky to construct efficiently implementable algorithms base on the dual-adversary solutions obtained from the learning graphs.

Thus it appears that even though efficient algorithms may exist for small values of $k$, the situation is not very pleasant for large $k$, especially $k=\Omega(n)$ --- the learning graph idea may not work (even if the corresponding algorithm could be implemented in a time-efficient manner) and the quantum walk algorithm uses $\Omega(k)$ space. Our algorithm addresses this concern and is specifically designed to use $\tilde{O}(1)$ qubits; as an added benefit, it works for any $k$.

\begin{lemma}\label{lemma:kd-ub}
There exists a bounded-error algorithm for \kdist, for any $k \in [n]$, that uses  $O(\tfrac{n^{3/2}}{\sqrt{k}}\log(\tfrac{1}{\delta \cdot k}))$ queries and $O\big((\log(m) + \log(n))\log(\frac{1}{\delta \cdot k})\big)$ qubits.
\end{lemma}

This algorithm has two attractive features. First is that it improves upon the algorithm proposed by Ambainis for $k \ge 4$ when we require that $\tilde{O}(1)$ space be used, and secondly its query complexity does not increase with $k$.

There have been separate attempts to design algorithms for specific values of $k$. For example, for $k=3$ Belovs designed a slightly different algorithm compared to the above~\cite{Belovs2014ApplicationsAlgorithms} and Childs et al.~\cite{Childs2013AUpdates} gave a random walk based algorithm both of which uses $O(n^{5/7})$ queries and $O(n^{5/7})$ space. These algorithm improved upon the $O(n^{3/2})$-query algorithm proposed earlier by Ambainis~\cite{Ambainis2007QuantumDistinctness}. Our algorithm provides an alternative that matches the query complexity of the latter and can come in handy when a small number of qubits are available.

For $k$ that is large, e.g. $\Omega(n)$, the query complexity of Ambainis' algorithm is exponential in $n$ and that of ours is $O(n^{3/2})$. Montanaro used a reduction from the \countdec problem~\cite{Nayak1999QuantumStatistics} to prove a lower bound of $\Omega(n)$ queries for $k=\Omega(n)$ --- of course, assuming unrestricted space~\cite{Montanaro2016TheMoments}. Our algorithm matches this lower bound, but with only $\tilde{O}(1)$ space.

\subsection{Upper bounds for the \gkd problem}

The \gkd problem was introduced by Montanaro~\cite[Sec~2.3]{Montanaro2016TheMoments} as a generalization of the \kdist problem to solve the \Finf problem; we modified the ``gap'' therein to additive to suit the results of this paper.

\begin{problem}[$\Delta$-\gkd]
This is the same as the \kdist problem along with a promise that either there exists a set of $k$ distinct indices with identical values or no value appears more than $k-\Delta$ times.
\end{problem}

Montanaro observed that this problem can be reduced to \Finf estimation and vice-versa with a $\log(n)$ overhead for binary search; however, he left open an algorithm or the query complexity of this problem. We are able to design a constant space algorithm by reducing it to our \prob problem. Our results are summarised in Table~\ref{table:summary-kdist}.

\begin{lemma}\label{lemma:dgkd-ub}
There is a quantum algorithm to solve the \dgkd problem that makes $\tilde{O}(\tfrac{n^{3/2}}{\Delta\sqrt{k}})$ queries and uses $O\big((\log(m)+\log(n))\log(\frac{1}{\delta \cdot k})\big)$ qubits.
\end{lemma}



\subsection{Upper bounds for \Finf}


The \Finf problem is a special case of the \pmaxprob problem on a finite array.

\begin{problem}[\Finf]
Given an oracle to query an $n$-sized array $A$ with values in $\{1, \ldots, m\}$, compute the frequency of the most frequent element, also known as the modal frequency.
\end{problem}

Li et al.~\cite{Li2019QuantumEstimation} studied this problem in the context of min-entropy of an array.
%
They reduced the problem of \minent estimation (of an $m$-valued array with additive error $\epsilon \in (0,1)$) to that of \kdist with $k=\lceil \frac{16\log(m)}{\epsilon^2} \rceil$. However they did not proceed further and made the remark that ``{\tt Existing quantum algorithms for the k-distinctness problem \ldots do not behave well for super-constant $k$s.}''. Indeed, it is possible to run the quantum-walk based algorithm for \kdist~\cite{Ambainis2007QuantumDistinctness} and thereby solve \Finf estimation; this turns out to be not very effective with $O(n)$ query complexity and $O(n)$ space complexity. (See Appendix~\ref{sec:li_wu_compare} for a rough analysis.)

Instead, we reduce the \Finf problem to that of \prob and obtain a $\tilde{O}(1)$-space algorithm to estimate the modal frequency with additive error. Montanaro proposed two methods to accurately compute the modal frequency, one of which closely matches the complexities of our proposed algorithm but our approach has a lower query complexity when $\epsilon=poly(1/n)$. The results are summarised in Table~\ref{table:summary-finf-ub}.

\begin{lemma}\label{lemma:finf-ub} There is a quantum algorithm to estimate \Finf with \dots
\begin{itemize}
    \item additive accuracy $\epsilon$ using  $O\big((\log(m) + \log(\frac{n}{\epsilon}))\log(\frac{n}{\delta\epsilon})\big)$ qubits and $\tilde{O}\left((\tfrac{n}{\epsilon})^{3/2} \log\tfrac{n}{\epsilon} \right)$ queries.
    \item relative accuracy $\epsilon$ using $O\big((\log(\frac{m}{\epsilon})+\log(n))\log(\frac{n}{\delta\epsilon})\big)$ qubits and $\tilde{O}(\frac{m^{3/2}}{\epsilon})$ queries. 
\end{itemize}
\end{lemma}

\paragraph{Heavy hitters:} A discrete version of the \prob problem has been studied as ``heavy hitters'' in the streaming domain, in which items (of an $n$-sized array) are given to an algorithm one by one, and the algorithm has to identify all items with frequency above a certain threshold, say $\tau n$. Since their objective was to return a list of items, naturally they used more than $\tilde{O}(1)$ space; further, even though they employed randomized techniques like sampling and hashing, they processed all items (query complexity is $O(n)$)~\cite{MankuMotwani,CMSketch,CountSketch}. The space required for all such algorithms are $\tilde{O}(\tfrac{1}{\epsilon})$ where $\epsilon$ indicates the permissible error during estimation of frequencies. Our approach decides if there is any heavy hitter, and if there are any, then samples from them; it makes use of only $\tilde{O}\big({\log\tfrac{1}{\epsilon}}\big)$ qubits.

A key feature of our algorithms is $o(1)$ queries to $D$. $o(1)$-query classical algorithms are possible if only sublinear samples are drawn. Valiant and Valiant showed that $O(\tfrac{m}{\epsilon^2\log m})$ samples are sufficient to construct an approximate histogram of $D$ (with support at most $m$) with additive ``error'' $\epsilon$~\cite{ValiantValiant}, and further showed that $\Omega(\tfrac{m}{\epsilon \log m})$ samples are necessary to compute some simple properties of $D$, such as Shannon entropy. It was not immediately clear to us if their lower bound extends to heavy hitters (or even the presence of heavy hitters); however, their approximate histogram can surely be used to identify them. Our quantum algorithm has a lower query complexity $\tilde{O}(\tfrac{1}{\epsilon^{3/2}})$.

\subsection{Non-linearity estimation of a Boolean function}

Non-linearity of a function $f$ is defined in terms of the largest absolute-value of its Walsh-Hadamard coefficient~\cite{Bera2021QuantumEstimation}: $\eta(f) = \tfrac{1}{2} - \tfrac{1}{2} \fmax$ where $\fmax = \max_x |\hat{f}(x)|$. Since the output state of the Deutsch-Jozsa circuit is $\sum_x \hat{f}(x)\ket{x}$, i.e., the probability of observing $\ket{x}$ is $\hat{f}(x)^2$, it immediately follows that we can utilize the \pmaxprob algorithm (that in itself uses \prob) to estimate $\fmax^2$, and hence, non-linearity, with additive inaccuracy. However, instead of \prob we can use \amprob and then use the same binary search strategy as \pmaxprob to estimate $\fmax$ instead of $\fmax^2$. This reduces the number of queries since the complexity of the binary-search based \pmaxprob algorithm depends upon $\pmax$ itself, and further a larger inaccuracy can be tolerated (to estimate $\fmax$ within $\pm \lambda$, it now suffices to call \amprob with inaccuracy $\lambda$, instead of calling $\prob$ with inaccuracy $\lambda^2$). This leads to a quadratic improvement in the query complexity in form of $\Ot(\frac{1}{\lambda\fmax})$. Details can be found in Appendix~\ref{sec:non-lin-est}.

\begin{lemma}
    \label{lemma:non-lin}
    Given a Boolean function $f:\{0,1\}^n\xrightarrow{}\{0,1\}$ as an oracle, an accuracy parameter $\lambda$ and an error parameter $\delta$, there exists an algorithm that returns an estimate $\tilde{\eta}_f$ such that $|\eta_f - \tilde{\eta}_f|\le \lambda$ with probability at least $1-\delta$ using $O(\frac{1}{\lambda\hat{f}_{max}}\log(\frac{1}{\lambda})\log(\frac{1}{\delta\hat{f}_{max}}))$ queries to the oracle of $f$. 
\end{lemma}

\newpage

\newpage
\bibliographystyle{plain}
\bibliography{references}

\newpage

\appendix

\section{Amplitude amplification, amplitude estimation and majority}\label{appendix:ampest-and-amp}

\newcommand{\MAJ}{\mathtt{MAJ}}

In this section, we present details on the quantum amplitude estimation and amplitude amplification subroutines that are used as part of our algorithms. We also explain the $\MAJ$ operator.

\subsection{Amplitude amplification}

The amplitude amplification algorithm (AA) is a generalization of the novel Grover's algorithm.
Given an $n$-qubit algorithm $A$ that outputs the state $\ket{\phi}=\sum_k\alpha_k\ket{k}$ on $\ket{0^n}$ and a set of basis states $G=\{\ket{a}\}$ of interest, the goal of the amplitude amplification algorithm is to amplify the amplitude $\alpha_a$ corresponding to the basis state $\ket{a}$ for all $\ket{a}\in G$ such that the probability that the final measurement output belongs to $G$ is close to 1.
In the most general setting, one is given access to the set $G$ via an oracle $O_G$ that marks all the states $\ket{a}\in G$ in any given state $\ket{\phi}$; i.e., $O_G$ acts as
$$O_G \sum_k\alpha_k\ket{k}\ket{0} \xrightarrow{}\sum_{a\notin G}\alpha_a\ket{a}\ket{0} + \sum_{a\in G}\alpha_a\ket{a}\ket{1}.$$

Now, for any $G$, any state $\ket{\phi} = \sum_k\alpha_k\ket{k}$ can be written as 
$$\ket{\phi} = \sum_k\alpha_k\ket{k} = \sin(\theta)\ket{\nu} + \cos(\theta)\ket{\overline{\nu}}$$ where $\sin(\theta) = \sqrt{\sum_{a\in G}|\alpha_a|^2}$, $\ket{\nu} = \frac{\sum_{a\in G}\alpha_a\ket{a}}{\sqrt{\sum_{a\in G}|\alpha_a|^2}}$ and $\ket{\overline{\nu}} = \frac{\sum_{a\notin G}\alpha_a\ket{a}}{\sqrt{\sum_{a\notin G}|\alpha_a|^2}}$.
Notice that the states $\ket{\nu}$ and $\ket{\overline{\nu}}$ are normalized and are orthogonal to each other.
The action of the amplitude amplification algorithm can then be given as
$$AA\Big(\sum_k\alpha_k\ket{k}\ket{0}\Big) = AA\big(\sin(\theta)\ket{\nu} + \cos(\theta)\ket{\overline{\nu}}\big)\ket{0} \xrightarrow{} \sqrt{(1-\beta)}\ket{\nu}\ket{1} + \sqrt{\beta}\ket{\overline{\nu}}\ket{0}$$ where $\beta$ satisfies $|\beta| < \delta$ and $\delta$ is the desired error probability.
This implies that on measuring the final state of AA, the measurement outcome $\ket{a}$ belongs to $G$ with probability $|1-\beta|$ which is at least $1-\delta$.

\subsection{Quantum amplitude estimation (QAE)}
Consider a quantum circuit $A$ on $n$ qubits whose final state is $\ket{\psi}$ on input $\ket{0^n}$. Let $\ket{a}$ be some basis state (in the standard basis --- this can be easily generalized to any arbitrary basis).
Given an accuracy parameter $\epsilon\in (0,1)$, the amplitude estimation problem is to estimate the probability $p$ of observing  $\ket{a}$ upon measuring $\ket{\psi}$ in the standard basis, up to an additive accuracy $\epsilon$.

Brassard et al., in~\cite{brassard2002quantum}, proposed a quantum amplitude estimation circuit, which we call $AmpEst$, that acts on two registers of size $m$ and $n$ qubits and makes $2^m-1$ calls to controlled-$A$ to output an estimate $\tilde{p} \in [0,1]$ of $p$ that behaves as mentioned below.
\begin{theorem}\label{thm:amp_est}
    The amplitude estimation algorithm returns an estimate $\tilde{p}$ that has a confidence
    interval $|p-\tilde{p}| \le 2\pi k \frac{\sqrt{p(1-p)}}{2^m} +
    \pi^2 \frac{k^2}{2^{2m}}$ with probability at least $\frac{8}{\pi^2}$ if
    $k=1$ and with probability at least $1-\frac{1}{2(k-1)}$ if $k \ge 2$. It uses exactly $2^m-1$ evaluations of the oracle. If
    $p=0$ or 1 then $\tilde{p}=p$ with certainty.
\end{theorem}

The following corollary is obtained directly from the above theorem.
\begin{corollary}\label{cor:amp_est_our_form}
    The amplitude estimation algorithm returns an estimate $\tilde{p}$ that has a confidence
    interval $|p-\tilde{p}| \le \frac{1}{2^q}$ with probability at least $\frac{8}{\pi^2}$ using $q+3$ qubits and $2^{q+3}-1$ queries. If
    $p=0$ or 1 then $\tilde{p}=p$ with certainty.
\end{corollary}
\begin{proof}
Set $k=1$ in Theorem~\ref{thm:amp_est}. Since $p\le 1$, we get $\sqrt{p(1-p)} \le \frac{1}{2}$. Then we have
$$2\pi k \frac{\sqrt{p(1-p)}}{2^m} + \pi^2 \frac{k^2}{2^{2m}} \le 2\pi\frac{1}{2\cdot 2^m} + \pi^2\frac{1}{2^{2m}} \le \frac{\pi}{2^m} + \frac{\pi^2}{2^{2m}}\le \frac{2\pi}{2^m} \le \frac{8}{2^m} = \frac{1}{2^{m-3}}.$$
The last inequality follows from the fact that $\frac{\pi}{2^m} < 1$ (which is true when $m \ge 2$). Now, set $m=q+3$ to prove the corollary.
\end{proof}


Now, let $p_a$ be the probability of obtaining the basis state $\ket{a}$ on measuring the state $\ket{\psi}$. The amplitude estimation circuit referred to above uses an oracle, denoted $O_a$ to mark the ``good state'' $\ket{a}$, and involves measuring the output of the $AmpEst$ circuit in the standard basis; actually, it suffices to only measure the first register. We can summarise the behaviour of the $AmpEst$ circuit (without the final measurement) in the following lemma.
\begin{lemma}
    Given an oracle $O_x$ that marks $\ket{x}$ in some state $\ket{\psi}$, $AmpEst$ on an input state $\ket{\psi}\ket{0^m}$ generates the following state.
    $$AmpEst \ket{\psi}\ket{0^m} \xrightarrow{} \beta_{x,s}\ket{\psi}\ket{\hat{p_x}} + \beta_{x,\overline{s}}\ket{\psi}\ket{E_x}$$
    where $|\beta_{x,s}|^2$, the probability of obtaining the good estimate, is at least $\frac{8}{\pi^2}$, and $\ket{\hat{p}_x}$ is an $m$-qubit normalized state of the form $\ket{\hat{p}_x} = \gamma_{+}\ket{\hat{p}_{x,+}} + \gamma_{-}\ket{\hat{p}_{x,-}}$ such that for $p\in \{\hat{p}_{x,+}, \hat{p}_{x,-}\} = \pmset_{p_x}~(say)$, $\sin^2(\pi\frac{p}{2^m})$ approximates $p_x$ up to $m-3$ bits of accuracy. Further, $\ket{E_x}$ is an $m$-qubit error state (normalized) such that any basis state in $\ket{E_x}$ corresponds to a bad estimate, i.e., we can express it as $\displaystyle\ket{E_x} = \sum_{t \in \{0,1\}^m}^{t\notin \pmset_{p_x}}\gamma_{t,x}\ket{t}$ in which $|\sin^2\left(\pi\tfrac{t}{2^m}\right) - p_x| > \tfrac{1}{2^{m-3}}$ for any $t \not\in \pmset_{p_x}$.
\end{lemma}

In an alternate setting where the oracle $O_x$ is not provided, $AmpEst$ can still be performed if the basis state $\ket{a}$ is provided --- one can construct a quantum circuit, say $EQ$, that takes as input $\ket{\phi}\ket{x}$ and marks the state $\ket{x}$ of the superposition state $\ket{\phi}$ as described in section~\ref{appendix:hd_algo_1}.
We name this extended-$AmpEst$ circuit as $\eqae$ which implements the following operation.

$$\eqae\big(\ket{x}\ket{\psi}\ket{0^m}\big) \xrightarrow{} \ket{x}\big(\beta_{x,s}\ket{\psi}\ket{\hat{p}_x} + \beta_{x,\overline{s}}\ket{\psi}\ket{E_x}\big)$$ where the notations are as defined above and the quantum circuit $EQ$ is used wherever the oracle $O_x$ was used in the previous setting.
In such a scenario, since $\eqae$ is a quantum circuit, we could replace the state $\ket{x}$ by a superposition $\sum_x\alpha_x\ket{x}$.
We then obtain the following.
\begin{corollary}
    Given an $EQ$ circuit, the $\eqae$ on an input state $\sum_x\alpha_x\ket{x}\ket{\psi}\ket{0^m}$ outputs a final state of the form
    $$\eqae\Big(\sum_x\alpha_x\ket{x}\ket{\psi}\ket{0^m}\Big) \xrightarrow{} \sum_x\alpha_x\beta_{x,s}\ket{x}\ket{\psi}\ket{\hat{p}_x} + \sum_x\alpha_x\beta_{x,\overline{s}}\ket{x}\ket{\psi}\ket{E_x}.$$
\end{corollary}
Notice that on measuring the first and the third registers of the output, with probability $|\alpha_x\beta_{x,s}|^2 \ge \frac{8}{\pi^2}|\alpha_x|^2$ we would obtain as measurement outcome a pair $\ket{a}\ket{x}$ where $\sin^2(\pi\frac{a}{2^m}) = \tilde{p}$ is within $\pm \frac{1}{2^{m-3}}$ of the probability $p_x$ of observing the basis state $\ket{x}$ when the state $\ket{\psi}$ is measured.
Observe in this setting that the subroutine essentially estimates the amplitude of all the basis states $\ket{x}$. However, with a single measurement we can obtain the information of at most one of the estimates. We will be using this in \etalgo.

  \subsection{$\MAJ$ operator}
Let $X_1 \ldots X_k$ be Bernoulli random variables with success probability $p > 1/2$. Let $Maj$ denote their majority value (that appears more than $k/2$ times). Using Hoeffding's bound\footnote{$\Pr[\sum X_i - E[\sum X_i] \ge t] \le \exp(-\frac{2t^2}{n})$}, it can be easily proved that $Maj$ has a success probability at least $1-\delta$, for any given $\delta$, if we choose $k \ge \tfrac{2p}{(p-1/2)^2} \ln \tfrac{1}{\delta}$. We require a quantum formulation of the same.

Suppose we have $k$ copies of the quantum state $\ket{\psi} = \ket{\psi_0} \ket{0} + \ket{\psi_1} \ket{1}$ in which we define ``success'' as observing $\ket{0}$ (without loss of generality) and $k$ is chosen as above. Let $p = \| \ket{\psi_0} \|^2$ denote the probability of success. Suppose we measure the final qubit after applying $(\mathbb{I}^k \otimes MAJ)$ in which the $MAJ$ operator acts on the second registers of each copy of $\ket{\psi}$. Then it is easy to show, essentially using the same analysis as above, that 
$$(\mathbb{I}^k \otimes MAJ) \ket{\psi}^{\otimes k} \ket{0} = \ket{\Gamma_0} \ket{0} + \ket{\Gamma_1} \ket{1}$$ in which $\| \ket{\Gamma_0} \|^2 \ge 1-\delta$.

The $\MAJ$ operator can be implemented without additional queries and with $poly(k)$ gates and $\log(k)$ qubits.

\section{Algorithms for \prob and \amprob problems}
\label{sec:entropy_test}
\label{appendix:hd_algo_1}

\subsection{Algorithm for \prob problem}
We design an algorithm for a promise version of \prob with additive error, which we refer to as \promiseprob.
For \prob we are given a quantum black-box $O_D$ such that
$\displaystyle O_D \ket{0^{log(m)}} \ket{0^a} = \sum_{x=0}^{m-1} \alpha_x \ket{x} \ket{\xi_x}$ in which $\{\ket{\xi_x} : x\in[m] \}$ are normalized states. Let $p_x=|\alpha_x|^2$ denote the probability of observing the first $\log(m)$ qubits in the standard-basis $\ket{x}$. The objective of \prob is to determine whether there exists any $x$ such that $p_x \ge \tau$ for any specified threshold $\tau \in (0,1)$ and the task of \pmaxprob is to compute $\max_x p_x$.




For this task we generalize {\tt QBoundFMax} from our earlier work on estimating non-linearity~\cite[Algorithm~3]{Bera2021QuantumEstimation}. The repurposing of that algorithm follows from three observations. First, {\tt QBoundFMax} identified whether there exists any basis state whose probability, upon observing the output of a Deutsch-Jozsa circuit, is larger than a threshold in a promised setting; however, no specific property of Deutsch-Jozsa circuit was being used. Secondly, amplitude estimation can be used to estimate $|\alpha_x|^2$ (with bounded error) in $\sum_x \alpha_x \ket{x} \ket{\xi_x}$ for any $x \in [n]$ by designing a sub-circuit on only the first $\log(m)$ qubits to identify ``good'' states (this sub-circuit was referred to as $EQ$ in {\tt QBoundFMax}).
Lastly, amplifying some states in a superposition retains their relative probabilities. These observations not only allow us to modify the {\tt QBoundFMax} algorithm for \prob, but also enable us to identify some $x$ such that $|\alpha_x|^2 \ge \tau$, along with an estimate of $|\alpha_x|^2$.

Our space-efficient algorithm for \promiseprob requires a few subroutines which we borrow from our earlier work on estimating non-linearity~\cite{Bera2021QuantumEstimation}.

\begin{description}
    \item[{\tt EQ$_m$}:] Given two computational basis states $\ket{x}$ and $\ket{y}$ each of $k$ qubits, {\tt EQ$_m$} checks if the {$m$-sized} prefix of $x$ and that of $y$ are equal. Mathematically, {\tt EQ$_m$}$\ket{x}\ket{y} = (-1)^{c}\ket{x}\ket{y}$ where $c=1$ if $x_i=y_i$ for all $i\in [m]$, and $c=0$ otherwise.
    \item[{\tt HD$_q$}:] When the target qubit is $\ket{0^q}$, and with a 
    $q-$bit string $y$ in the control register, {\tt HD} computes the absolute 
    difference of $y_{int}$ from $2^{q-1}$ and outputs it as a string where 
    $y_{int}$ is the integer corresponding to the string $y$. It can be represented 
    as ${\tt HD}_q \ket{y}\ket{b} = \ket{b\oplus\tilde{y}}\ket{y}$ where $y,b\in \{0,1\}^q$ 
    and $\tilde{y}$ is the bit string corresponding to the integer $\abs{2^{q-1} - y_{int}}$.
    Even though the operator {\tt HD} requires two registers, the second register 
    will always be in the state $\ket{0^q}$ and shall be reused by uncomputing (using
    $HD^\dagger$) after the {\tt CMP} gate. For all practical purposes, this operator 
    can be treated as the mapping $\ket{y} \mapsto \ket{\tilde{y}}$.
    \item[{\tt CMP}:] The ${\tt CMP}$ operator is defined as 
    ${\tt CMP} \ket{y_1}\ket{y_2}\ket{b} = \ket{y_1}\ket{y_2}\ket{b \oplus (y_2\le y_1)}$ 
    where $y_1, y_2\in \{0,1\}^n$ and $b\in \{0,1\}$. It simply checks if the integer corresponding to the basis state in the first register is at most that in the second register.
    \item[{\tt Cond-MAJ}:] The ${\tt Cond-MAJ}$ operator is defined as $\prod_x\big(\ket{x}\bra{x}\otimes MAJ\big)$ where $\ket{x}\bra{x}\otimes MAJ$ acts on computational basis states as $MAJ\ket{a_1}\cdots \ket{a_k}\ket{b} = \ket{a_1}\cdots \ket{a_k}\ket{b \oplus (\tilde{a} \ge k/2)}$ where $\tilde{a} = \sum_k a_k$ and $a_i,b \in \{0,1\}$.
\end{description}   

\begin{figure}
    \centering
    \includegraphics[width=\linewidth]{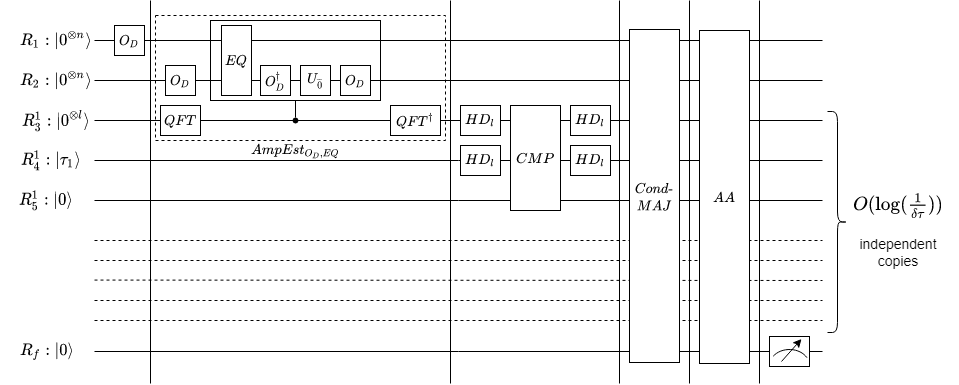}
    \caption{The quantum circuit corresponding to \etalgo. The stages are separated by dotted lines. $AmpEst$ is the circuit for amplitude estimation without the final measurement step.}
    \label{fig:hdalgo}
\end{figure}

\begin{algorithm}
    \caption{Algorithm \etalgo \label{algo:et_algo}}
    \begin{algorithmic}[1]
        \Require Oracle $O_D$ (with parameters $m$, $a$), threshold $\tau$, accuracy $\epsilon$ and error $\delta$.
        \State Set $r=\log(m)+a$, $\tau' = \tau - \frac{\epsilon}{8}$, $q = \lceil \log(\frac{1}{\epsilon}) \rceil +4$, $l= q+3$ and $c=\frac{1}{2(8/\pi^2 - 1/2)^2}$.
        \State Set $\tau_1 = \left\lfloor{\frac{2^l}{\pi}\sin^{-1}(\sqrt{\tau'})}\right\rfloor$
        \State Initialize $3$ registers $R_1R_2R_3$ as $\ket{0^\logn}\ket{0^\logn}\ket{\tau_1}$ and $c\ln(\frac{1}{\delta^2\tau^2})$ many independent copies of $R^k_4 R^k_5 = \ket{0^l}\ket{0}$ .
        The $3^{rd}$ register is on $l$ qubits.
        \State {\bf Stage 1:} Apply $O_D$ on $R_1$ and $R_2$.
        \On{$c\cdot \ln(\frac{1}{\delta^2\tau^2})$ many independent copies}
            \State {\bf Stage 2:} Apply quantum amplitude estimation sans measurement ($AmpEst$) on
            $O_D$ with $R_2$ as the input register, $R^k_4$ as the precision register and 
            $R_1$ is used to determine the ``good state''. $AmpEst$ is called with 
            error at most $1- \frac{8}{\pi^2}$ and additive accuracy $\frac{1}{2^q}$.\label{line:amp_est}
            \State {\bf Stage 3:} Use ${\tt HD_l}$ on $R_3$ and $R^k_4$
            individually.\label{line:half_dist}
            \State Use ${\tt CMP}$ on $R_3 = \ket{\tau_1}$ and $R^k_4$ as input registers 
            and $R^k_5$ as output register.\label{line:q_compare}
            \State Use ${\tt HD^{\dagger}_l}$ on $R_3$ and $R^k_4$
            individually.\label{line:half_dist_inv}
        \EndOn
        \State {\bf Stage 4:} For each basis state $\ket{x}$ in $R_1$, for $i=1 \ldots c\cdot\ln(\frac{1}{\delta^2\tau^2})$ compute the majority of the basis states of each $R^i_5$ register conditioned on the $R_1$ to be in $\ket{x}$, and store the result in $R_{f}$.
        \State {\bf Stage 5:} Apply Amplitude Amplification 
        (AA) $O(\frac{1}{\sqrt{\tau}})$ times on $R_f$ with error at most $\delta/2$ using $\ket{1}$ as the good state
        and measure $R_f$ as $out$.
        \State If $out = \ket{1}$ \Return{\tt TRUE} else \Return{\tt FALSE}\label{line:finish}
    \end{algorithmic}
\end{algorithm}

The algorithm for \promiseprob with additive accuracy is presented as \etalgo in Algorithm~\ref{algo:et_algo}.
The quantum circuit of the algorithm is illustrated in Figure~\ref{fig:hdalgo}.
%
Its operation can be explained in stages.
For convenience, let us call the set $G = \{\ket{z} : z\in [m], p_z\ge \tau\}$ 
as the `\textit{good}' set and its elements as the `\textit{good}' states.
In the first stage, we initialize the registers $R_1R_2R_3$ in the state
$\ket{0^\logn}\ket{0^\logn}\ket{\tau_1}$.
We then apply the oracle $O_D$ on $R_1$ and $R2$ to obtain the state 
of $R_1$ and $R2$ as $\sum_{x\in [m]} \alpha_x\ket{x}\ket{\xi_x}$.
Let $\ket{x,\xi_x}$ denote the state $\ket{x}\ket{\xi_x}$.

In stage two, we initialize $c\cdot \ln(\frac{1}{\delta^2\tau^2})$ copies of the registers $R_4^k R_5^k$ in the state $\ket{0^l}\ket{0}$.
For all $k=1\cdots c\cdot\ln(\frac{1}{\delta^2\tau^2})$, we then apply amplitude estimation collectively on the registers $R_1, R_2$ 
and $R_4^k$ in a way that for every basis state $\ket{z}$ in the first $\log(m)$ qubits of $R_1$, a string $a_z$ is output on 
$R_4^k$ such that $\sin^2{(\frac{a_z \pi}{2^q})} = \tilde{p_z}(say) \in  [p_z-\frac{1}{2^q}, p_z+\frac{1}{2^q}]$ with probability at least $8/\pi^2$.

Stage three is essentially about filtering out the \textit{good} states. 
We use the subroutines \hdq~and \cmp~to perform the filtering 
and marking all the \textit{good} states $\ket{z}$ by flipping 
the state of $R_5$ to $\ket{1}$ for such states.
So, the state in the circuit after stage three is $\ket{\psi_3} = 
\sum_{x\in [n]}\alpha_x \ket{x,\xi_x} \ket{\phi}\ket{a_x}\ket{\tau_1}\ket{\tilde{p}_x\ge \tau_1}$.
Notice that the probability of measuring $R_5$ as $\ket{1}$ in $\ket{\psi_3}$ is either $0$ or is lower bounded by $\tau$ due to the promise.

In stage four, for each basis state $\ket{x}$, we perform a conditional majority over all $R_5^k$ registers conditioned on the $R_1$ being $\ket{x}$ and store the result in a new register $R_f$. 
This stage ensures that the error caused due to amplitude estimation does not amplify to more than $\delta$ during the amplitude amplification in stage five.

Finally at stage five, we use the amplitude amplification to
amplify the probability of obtaining the state $\ket{1}$ in $R_f$.
Now for any $x$ that is marked, we have $\tilde{p}_x \ge \tau_1$.
If the probability of observing $\ket{1}$ in $R_f$ is non-zero, then since we have a lower bound on that probability, we have an upper bound on the number of amplifications needed to observe the state 
$\ket{1}$ in $R_f$ with high probability.

The above exposition is a simplified explanation of the algorithm that does not take into account errors and inaccuracies, especially those arising from amplitude estimating and interfering with amplitude amplification. 
The detailed proof of correctness and query complexity of the algorithm is discussed in the proof of Lemma~\ref{lemma:etalgo_correct} below.

Algorithm \etalgo contains an easter egg. When the output register is observed in the state $\ket{1}$, for majority of $k \in \{1, 2, \ldots c\log(\tfrac{1}{\delta^2 \tau^2})\}$, $R^k_5$ would be in $\ket{1}$ with high probability; the index register $R_1$ would contain some superposition of all good $x$'s.

The algorithm for \promiseprob with additive error can be used to solve \promiseprob with relative error $\epsilon_r$ by setting $\epsilon=\epsilon_r \tau$ 



\label{appendix:hd_algo}

\etalgolemma*


\begin{proof}
Before we provide the correctness of the algorithm we introduce a few propositions that will be useful in proving the correctness of the algorithm.

\begin{proposition}[Proposition~4.1,\cite{Bera2021QuantumEstimation}]\label{lemma:sin_inequality}
For any two angles $\theta_1, \theta_2 \in [0,\pi]$, $$\sin{\theta_1}\le \sin{\theta_2} \iff \sin^2{\theta_1}\le \sin^2{\theta_2} \iff \abs{\frac{\pi}{2}-\theta_1} \ge \abs{\frac{\pi}{2}-\theta_2}.$$
\end{proposition}


\begin{proposition}[Proposition~4.3,\cite{Bera2021QuantumEstimation}]
\label{prop:tau'_tau1_rel}
The constants $\tau'$ and $\tau_1$ in \etalgo satisfy $0\le \tau' - \frac{2\pi}{2^l}\le \sin^2(\frac{\pi \tau_1}{2^l})$.
\end{proposition}

We now analyse the algorithm. Recall that $\displaystyle O_D \ket{0^{log(m)}} \ket{0^a} = \sum_{x=0}^{m-1} \alpha_x \ket{x} \ket{\xi_x}$ which we denote $\ket{\phi}$, and $p_x = |\alpha_x|^2$.

\paragraph*{\bf Stage-1:} Consider the registers $R_1R_2R_3$ along with one of the $c\ln(\tfrac{1}{\delta^2\tau^2})$ independent copies and neglect the superscript on the registers. The state of the circuit after stage-1, just before amplitude estimation, is 
$$\ket{\psi_1} = \ket{\phi} \ket{\phi} \ket{\tau_1}\ket{0^q}\ket{0} = \sum_{x\in [m]}\alpha_x\ket{x,\xi_x}\ket{\phi}\ket{\tau_1}\ket{0^q}\ket{0}.$$

\paragraph*{\bf Stage-2:} After the amplitude estimation step, we obtain a state of the form $$\ket{\psi_2} = \sum_{x\in [m]}\alpha_x\ket{x,\xi_x}\ket{\phi}\ket{\tau_1}\Big(\beta_{x,s}\ket{a_x}+\beta_{x,\overline{s}}\ket{E_x}\Big)\ket{0}$$ where 
$\ket{a_x}$ is a normalized state of the form $\ket{a_x} = \gamma_{+}\ket{a_{x,+}} + \gamma_{-}\ket{a_{x,-}}$ that on measurement outputs $a \in \{a_{x,+}, a_{x,+}\}$ which is
an $l$-bit string that behaves as $\displaystyle\left| \sin^2\left(\frac{a\pi}{2^l}\right) - p_x \right| \le \tfrac{1}{2^q}$.
We denote the set $\{a_{x,+}, a_{x,-}\}$ by $\pmset_{a_x}$.

\paragraph*{\bf Stage-3:} 
Notice that stage 3 affects only the registers $R_3, R_4$ and $R_5$.
For any computational basis state $\ket{u}$ and $\ket{v}$, the transformation of a state of the form $\ket{u}\ket{v}\ket{0}$ due to stage 3 can be given as \begin{equation}\label{eqn:stage3}
    \ket{u}\ket{v}\ket{0} \xrightarrow{}\ket{u}\ket{v}\ket{\mathbb{I}\{u\ge v\}} \mbox{ where ~}\mathbb{I}\{u\ge v\}=1 \mbox{ if $u\ge v$ and $0$ else}.
\end{equation}
The reason for indicating ``$u \ge v$'' as 1 and not the other way around is due to the reversal of the direction of the inequality in Proposition~\ref{lemma:sin_inequality}.  Then, stage 3 transforms the state $\ket{\psi_2}=$
\begin{align*}
    &= \sum_{x\in [m]}\alpha_x\ket{x,\xi_x}\ket{\phi}\ket{\tau_1}\Big(\beta_{x,s}\ket{a_x}+\beta_{x,\overline{s}}\ket{E_x}\Big)\ket{0}\\
    &= \sum_{x\in [m]}\alpha_x\ket{x,\xi_x}\ket{\phi}\ket{\tau_1}\Big(\beta_{x,s}\gamma_{+}\ket{a_{x,+}}\ket{0} + \beta_{x,s}\gamma_{-}\ket{a_{x,-}}\ket{0} +\beta_{x,\overline{s}} \sum_{a\notin \pmset_{a_x}}\gamma_{x,a}\ket{a}\ket{0}\Big)\\
\end{align*}
to the state $\ket{\psi_3}=$
\begin{align*}
    &\sum_{x\in [m]}\alpha_x\ket{x,\xi_x}\ket{\phi}\ket{\tau_1}\Big(\beta_{x,s}\gamma_{+}\ket{a_{x,+}}\ket{\mathbb{I}\{a_x\le \tau_1\}} + \beta_{x,s}\gamma_{-}\ket{a_{x,-}}\ket{\mathbb{I}\{a_x\le \tau_1\}} \\
    & \cuquad + \beta_{x,\overline{s}} \sum_{a\notin \pmset_{a_x}}\gamma_{x,a}\ket{a}\ket{\mathbb{I}\{a\le \tau_1\}}\Big)\\
    &= \sum_{x\in [m]}\alpha_x\ket{x,\xi_x}\ket{\phi}\ket{\tau_1}\Bigg[\beta_{x,s} \Bigg\{ \gamma_{+}\ket{a_{x,+}}\ket{\mathbb{I}\{a_x\le \tau_1\}} + \gamma_{-}\ket{a_{x,-}}\ket{\mathbb{I}\{a_x\le \tau_1\}} \Bigg\} \\
    &\cuquad + \beta_{x,\overline{s}} \Bigg\{ \sum_{\substack{a\notin \pmset_{a_x} \\ a>\tau_1}}\gamma_{x,a}\ket{a}\ket{0} + \sum_{\substack{a\notin \pmset_{a_x} \\ a\le\tau_1}}\gamma_{x,a}\ket{a}\ket{1} \Bigg\}\Bigg]\tag{Eqn.~\ref{eqn:stage3}}
\end{align*}

We will analyse the states $\ket{\mathbb{I}\{a_{x\pm}\le \tau_1\}}$ by considering two types of index $x \in [m]$.

\paragraph*{Scenario (i):} $x$ be such that $p_x < \tau-\epsilon$. Now, any computational basis state $a$ in $\ket{a_x}$ will be such that $\tilde{p}_x (say) = \sin^2(\frac{a \pi}{2^l}) \in [p_x -\tfrac{1}{2^q}, p_x + \tfrac{1}{2^q}]$. Therefore, $\tilde{p}_x \le p_x + \tfrac{1}{2^q} < \tau - \epsilon + \tfrac{1}{2^q}$ and since $q$ was chosen such that $2^q \ge \frac{16}{\epsilon}$, $\tilde{p}_x < \tau - \frac{7\epsilon}{8}$.

Since we have $2^l > 2^q \ge \frac{16}{\epsilon}$, we get that $\frac{2\pi}{2^l} < \frac{2\pi\epsilon}{16} < \frac{6\epsilon}{8}$.
Using this, we have $\tau-\frac{7\epsilon}{8} = \tau' - \frac{6\epsilon}{8} < \tau'-\frac{2\pi}{2^l} < \sin^2(\frac{\pi \tau_1}{2^l})$ using Proposition~\ref{prop:tau'_tau1_rel}. Since, $\tilde{p}_x < \tau-\frac{7\epsilon}{8}$, we have $\tilde{p}_x = \sin^2(\frac{a\pi}{2^l}) < \tau - \frac{7\epsilon}{8} < \sin^2(\frac{\pi \tau_1}{2^l})$.

Now on applying \texttt{HD$_l$} on $R_3$ and $R_4$, 
we obtain $\ket{\hat{\tau}_1}$ and $\ket{\hat{a}}$ respectively in $R_3$ and $R_4$ such that $\hat{a} = |2^{l-1} - a|$ and 
$\hat{\tau}_1 = |2^{l-1} - \tau_1|$.
Using Proposition~\ref{lemma:sin_inequality} on the fact that $\tilde{p}_x  = \sin^2(\frac{a\pi}{2^l}) < \sin^2(\frac{\pi \tau_1}{2^l})$ 
we get $\hat{a}  = |2^{l-1} - a| > \hat{\tau}_1  = |2^{l-1} - \tau_1|$.

Since 
$\hat{a} > \hat{\tau}_1$ corresponding to any $\ket{a} \in \{ \ket{a_{x,-}}, \ket{a_{x,+}} \}$, after using \cmp~on $R_3$ and $R_4$, we get in $R_5$ the state $\ket{\mathbb{I}\{a_{x,-}\le \tau_1\}}=\ket{\mathbb{I}\{a_{x,+}\le \tau_1\}}=\ket{0}$. 

\paragraph*{Scenario (ii):} $z$ be such that $p_z \ge \tau$. 
 Now, any computational basis state $a$ in $\ket{a_z}$ will be such that $\tilde{p}_z (say) = \sin^2(\frac{a \pi}{2^l}) \in [p_z -\tfrac{1}{2^q}, p_z + \tfrac{1}{2^q}]$. Therefore, $\tilde{p}_z \ge p_z - \tfrac{1}{2^q} > \tau - \tfrac{1}{2^q}$ and since $q$ was chosen such that $2^q \ge \frac{16}{\epsilon}$, $\tilde{p}_z = \sin^2(\frac{a \pi}{2^l}) > \tau - \frac{\epsilon}{8} = \tau'$. 
 
This gives us $\sin(\frac{a\pi}{2^l}) > \sqrt{\tau'}$ since $\frac{a\pi}{2^l} \in [0,\pi]$. Furthermore, since $\tau_1$ is an integer in $[0,2^{l}-1]$ and $\tau_1 = \left\lfloor\frac{2^l}{\pi}\sin^{-1}(\sqrt{\tau'})\right\rfloor \le \frac{2^l}{\pi}\sin^{-1}(\sqrt{\tau'})$, we get $\sqrt{\tau'} \ge \sin(\frac{\tau_1\pi}{2^l})$. Combining both the inequalities above, we get $\sin(\frac{a\pi}{2^l}) > \sin(\frac{\tau_1\pi}{2^l})$.

Now, on applying \texttt{HD$_l$} on $R_3$ and $R_4$, 
we obtain $\ket{\hat{\tau}_1}$ and $\ket{\hat{a}}$ respectively in $R_3$ and $R_4$ such that $\hat{a} = |2^{l-1} - a|$ and 
$\hat{\tau}_1 = |2^{l-1} - \tau_1|$. 
Using Proposition~\ref{lemma:sin_inequality} on the fact that $\sin(\frac{a\pi}{2^l}) > \sin(\frac{\tau_1\pi}{2^l})$, we get $\hat{a} < \hat{\tau}_1$.

As above, since $\hat{a} < \hat{\tau}_1$ corresponding to any $\ket{a} \in \{ \ket{a_{x,-}}, \ket{a_{x,+}} \}$, after using \cmp~on $R_3$ and $R_4$, we get in $R_5$ the state $\ket{\mathbb{I}\{a_{x,-}\le \tau_1\}}=\ket{\mathbb{I}\{a_{x,+}\le \tau_1\}}=\ket{1}$.


\paragraph*{\bf Stage-4 and Stage-5:}
It is evident from the above analysis that $R_5$ is correctly set to $\ket{0}$ or $\ket{1}$ for $x$ such that $p_x < \tau - \epsilon$ or $p_x \ge \tau$, respectively, however only with certain probability. In fact, amplitude estimation will not succeed with some probability, and will yield some $a \not\in S_{a_x}$ in $R_4$ some of which may produce erroneous results in $R_5$ after comparison with $\tau_1$ in $R_3$. We need to pin down the probability of error to analyse this stage. For this, we consider the two scenarios corresponding to the promises of \promiseprob.

\paragraph*{Case (i):} Consider the case when for all $x\in [m]$, $p_x < \tau-\epsilon$.
Then, that state after stage 3 can be written as
\begin{align*}
    \ket{\psi_4} &= \sum_{x\in [m]}\alpha_x\ket{x}\ket{\phi}\ket{\tau_1}\Bigg[\beta_{x,s}\ket{a_x}\ket{0}\\
    &\cuquad + \beta_{x,\overline{s}} \Bigg\{ \sum_{\substack{a\notin \pmset_{a_x} \\ a>\tau_1}}\gamma_{x,a}\ket{a}\ket{0} + \sum_{\substack{a\notin \pmset_{a_x} \\ a\le\tau_1}}\gamma_{x,a}\ket{a}\ket{1} \Bigg\}\Bigg]\\
\end{align*}
Recall that $\beta_{x,\overline{s}} \le 1 - \frac{8}{\pi^2} < 0.2$ for all $x$. Therefore, on measuring $R_5$, the probability of obtaining $\ket{1}$ (false positive) can be given as 
$$Pr\big[R_5=\ket{1}\big] = \sum_{x\in[m]} |\alpha_x\beta_{x,\overline{s}}|^2 \sum_{\substack{a\notin \pmset_{a_x}\\ a\le \tau_1 }}|\gamma_{x,a}|^2 \le 0.2 \sum_{x\in[m]} |\alpha_x|^2 \sum_{\substack{a\notin \pmset_{a_x}\\ a\le \tau_1 }}|\gamma_{x,a}|^2 \le \sum_{x\in[m]} |\alpha_x|^2 \cdot 1 \le 0.2 .$$

At this point we perform the conditional majority operator on $c\log(\frac{1}{\delta^2 \tau^2})$ independent copies of $R_5$ conditioned on $R_1$ being in the basis state $\ket{x}$ for each $x\in \{0,1\}^n$. 
Then using Hoeffding's inequality, the following relation is straight forward for each $x\in\{0,1\}^n$:
$$Pr\Big[R_f=\ket{1}|R_1=\ket{x}\Big] \le \delta^2\tau^2.$$
As this relation is true for each $x\in\{0,1\}^n$, we have that $Pr\Big[R_f=\ket{1}\Big] \le \delta^2\tau^2$.

In stage 5 we perform any of the amplitude amplification algorithms that can operate using a lower bound on the success probability~\cite{brassard2002quantum}. Since there are many such methods, we avoid choosing any specific one; however, all of them will involve some $m$ iterations where $m=O(\frac{t_\delta}{\sqrt{0.8\tau}})$ for some suitable $t$. 
We will now show that even after amplitude amplification with $m$ iterations, the probability of false positive will be at most $\delta$. 

Notice that $O(\frac{t_\delta}{{\delta \cdot \tau}})$ iterations are required to amplify a minimum probability of $\delta^2\tau^2$ to $\delta$. But $\frac{t_\delta}{{\delta\cdot \tau}} \gg \frac{t_\delta}{\sqrt{0.8\tau}}$; hence, even after amplifying with $m$ iterations, $R_f$ can be observed in the state $\ket{1}$ with probability less than $\delta$.

Hence, if $p_x < \tau-\epsilon$ for all $x\in [m]$, the probability of obtaining state $\ket{0}$ on measuring $R_f$ is at least $1-\delta$.

\paragraph*{Case (ii):}
In this case there exists some $z\in [m]$ such that $p_z \ge \tau$; in fact, let $G=\{z\in [m] : p_x\ge \tau\}$. Then the state after stage $3$ can be given as
\begin{align*}
    \ket{\psi_4} &= \sum_{x\in [m]}\alpha_x\ket{x}\ket{\phi}\ket{\tau_1}\Bigg[\beta_{x,s}\ket{a_x}\ket{\mathbb{I}\{a_x\le \tau_1\}}\\
    &\cuquad + \beta_{x,\overline{s}} \Bigg\{ \sum_{\substack{a\notin \pmset_{a_x} \\ a>\tau_1}}\gamma_{x,a}\ket{a}\ket{0} + \sum_{\substack{a\notin \pmset_{a_x} \\ a\le\tau_1}}\gamma_{x,a}\ket{a}\ket{1} \Bigg\}\Bigg]\\
    & = \sum_{x\in G}\alpha_x\ket{x}\ket{\phi}\ket{\tau_1}\Bigg[\beta_{x,s}\ket{a_x}\ket{1}\\
    &\cuquad +\beta_{x,\overline{s}} \Bigg\{ \sum_{\substack{a\notin \pmset_{a_x} \\ a>\tau_1}}\gamma_{x,a}\ket{a}\ket{0} + \sum_{\substack{a\notin \pmset_{a_x} \\ a\le\tau_1}}\gamma_{x,a}\ket{a}\ket{1} \Bigg\}\Bigg]\\
    & + \sum_{x\notin G}\alpha_x\ket{x}\ket{\phi}\Bigg[\beta_{x,s}\ket{a_x}\ket{\tau_1}\ket{0}\\
    &\cuquad +\beta_{x,\overline{s}} \Bigg\{ \sum_{\substack{a\notin \pmset_{a_x} \\ a>\tau_1}}\gamma_{x,a}\ket{a}\ket{0} + \sum_{\substack{a\notin \pmset_{a_x} \\ a\le\tau_1}}\gamma_{x,a}\ket{a}\ket{1} \Bigg\}\Bigg]
\end{align*}

Notice that in the above summation, we simply break the state $\ket{\psi_3}$ into two summands of which one contains the summation over all $x\in G$ and the other contains the summation over all $x\notin G$. 

Next, in stage-4, for every $x\in\{0,1\}^n$, conditioned on the register $R_1$ being in state $\ket{x}$, we perform a conditional majority over all the $R^k_5$ registers and store the output in $R_{f}$.
Then, using Hoeffding's bound as in case(i), we get that for any $x\in G$, $$Pr\Big[R_{f}=\ket{1}\Big|R_1=\ket{x}\Big] \ge 1-\delta^2\tau^2 \ge 1-\delta,$$ and for any $x\notin G$ we have, $$Pr\Big[R_{f}=\ket{1}\Big|R_1=\ket{x}\Big] \le \delta^2 \tau^2 < \delta.$$

Therefore, the overall probability of obtaining $\ket{1}$ in $R_{f}$ after stage-5 can be expressed as
$$Pr\Big[R_{f}=\ket{1}\Big] \ge \sum_{x\in G}|\alpha_x|^2\cdot (1-\delta) \ge \tau(1-\delta) \ge \tau/2$$ under the reasonable assumption that the target error probability $\delta < \tfrac{1}{2}$.

Now, we present the query complexity of the algorithm. It is obvious that the number of calls made by amplitude estimation with accuracy 
$\frac{1}{2^q}$ and error at most $1-\frac{8}{\pi^2}$ is $O(2^q) = O(\frac{1}{\epsilon})$. 
The subroutines \hdq and \cmp are query independent.
In total, we perform $\log(\frac{1}{\delta'}) = O(\log(\frac{1}{\delta\tau}))$ many independent estimates and comparisons in the worst case.
Again, computing the majority does not require any oracle queries.
In the final stage, we perform the amplitude amplification with $O(\frac{1}{\sqrt{\tau}})$ iterations.
Hence, the total number of oracle queries made by the algorithm is $O(\frac{1}{\epsilon\sqrt{\tau}}\log(\frac{1}{\delta\tau}))$ queries.

Finally, the number of qubits used in the algorithm is $2*r+\big((2*l+1)\cdot \log(\tfrac{1}{\delta\tau})\big) = O\big(\log(m)+\big((\log(\frac{1}{\epsilon})+a)\cdot \log(\tfrac{1}{\delta\tau})\big)\big)$.

\end{proof}

\subsection{Algorithm for \amprob problem}

We present the algorithm for \amprob problem as Algorithm~\ref{algo:amprob_algo}. The algorithm for \amprob differs from the \etalgo only at stages-1 and 2.

\begin{algorithm}
    \caption{Algorithm \highampalgo for \amprob problem \label{algo:amprob_algo}}
    \begin{algorithmic}[1]
        \Require Oracle $O_D$ (with parameters $m$, $a$), threshold $\tau$, accuracy $\epsilon$ and error $\delta$.
        \State Set $r=\log(m)+a$, $\tau' = \frac{1}{2}(1-\tau + \frac{\epsilon}{8})$, $q = \lceil \log(\frac{1}{\epsilon}) \rceil +4$, $l= q+3$ and $c=\frac{1}{2(8/\pi^2 - 1/2)^2}$.
        \State Set $\tau_1 = \left\lfloor{\frac{2^l}{\pi}\sin^{-1}(\sqrt{\tau'})}\right\rfloor$
        \State Initialize $4$ registers $R_1R_{21}R_{22}R_3$ as $\ket{0^\logn}\ket{0}\ket{0^{\logn}}\ket{\tau_1}$ and $c\ln(\frac{1}{\delta^2\tau^2})$ many independent copies of $R^k_4 R^k_5 = \ket{0^l}\ket{0}$ .
        The $3^{rd}$ register is on $l$ qubits.
        \State {\bf Stage 1:} Apply $O_D$ on $R_1$.
        \State For all $x$, controlled on $R1$ being in state $\ket{x}$, apply Hadamard test on $R_{21}R_{22}$ with $A_{\psi}=A_x$ and $A_{\phi}= O_D$ where $A_x\ket{0^n}=\ket{x}$.
        \On{$c\cdot \ln(\frac{1}{\delta^2\tau^2})$ many independent copies}
            \State {\bf Stage 2:} Apply simultaneous amplitude estimation ($SAE$) sans measurement with $R_{21}$ as the input register, $R^k_4$ as the precision register and $\ket{0}$ as the ``good state''. $SAE$ is called with
            error at most $1- \frac{8}{\pi^2}$ and additive accuracy $\frac{1}{2^q}$.\label{line:amp_est}
            \State {\bf Stage 3:} Use ${\tt HD_l}$ on $R_3$ and $R^k_4$
            individually.\label{line:half_dist}
            \State Use ${\tt CMP}$ on $R_3 = \ket{\tau_1}$ and $R^k_4$ as input registers 
            and $R^k_5$ as output register.\label{line:q_compare}
            \State Use ${\tt HD^{\dagger}_l}$ on $R_3$ and $R^k_4$
            individually.\label{line:half_dist_inv}
        \EndOn
        \State {\bf Stage 4:} For each basis state $\ket{x}$ in $R_1$, for $i=1 \ldots c\cdot\ln(\frac{1}{\delta^2\tau^2})$ compute the majority of the basis states of each $R^i_5$ register conditioned on the $R_1$ to be in $\ket{x}$, and store the result in $R_{f}$.
        \State {\bf Stage 5:} Apply Amplitude Amplification 
        (AA) $O(\frac{1}{\sqrt{\tau}})$ times on $R_f$ with error at most $\delta/2$ using $\ket{0}$ as the good state
        and measure $R_f$ as $out$.
        \State If $out = \ket{0}$ \Return{\tt TRUE} else \Return{\tt FALSE}\label{line:finish}
    \end{algorithmic}
\end{algorithm}

Before we prove the correctness of the algorithm, we establish the following proposition whose proof is straightforward.

\begin{proposition}
    \label{prop:fhat-tau-scale-change}
    For any $\alpha_x$, a threshold $\tau$ and some $\epsilon$,
    \begin{enumerate}
        \item $|\alpha_x|\ge \tau+2\epsilon \iff \frac{1}{2}\big(1-|\alpha_x|\big)\le \frac{1}{2}\big(1-\tau\big)-\epsilon$.
        \item $|\alpha_x|< \tau \iff \frac{1}{2}\big(1-|\alpha_x|\big)> \frac{1}{2}\big(1-\tau\big)$.
    \end{enumerate}
\end{proposition}

\begin{proof}[Proof of Algorithm~\ref{algo:amprob_algo}]
    We now analyse the algorithm stage by stage.

    \paragraph*{\bf Stage-1:} Consider the registers $R_1R_{21}R_{22}R_3$. 
    The state of these registers after stage-1 can be given as
    $$\ket{\psi_1'} = \frac{1}{2}\sum_{x\in [m]}\alpha_x\ket{x,\xi_x}\big(\ket{0}(\ket{x}+\ket{\phi}) + \ket{1}(\ket{x}-\ket{\phi})\big)$$ where $\ket{\phi}=\sum_x\alpha_x\ket{x,\xi_x}$.
    Given that the state in $R_1$ is $\ket{x,\xi_x}$, the probability of obtaining $\ket{0}$ in $R_{21}$ can then be given as $$Pr(\ket{0}_{R_{21}}) = \frac{1}{4}||\ket{x}+\ket{\phi}||^2 = \frac{1}{4}\big(2+2\bra{x}\ket{\phi}\big) = \frac{1}{2}(1+|\alpha_x|).$$
    Using this, $\ket{\psi_1'}$ can be given as
    $$\ket{\psi_1'} = \sum_{x\in[m]}\alpha_x\ket{x,\xi_x}\Big(\nu_{x0}\ket{0}\ket{\eta_{x0}} + \nu_{x1}\ket{1}\ket{\eta_{x1}}\Big) = \sum_x\alpha_x\ket{x,\xi_x}\ket{\nu_x}\text{~(say)}$$ for some normalized states $\ket{\eta_{x0}}$ and $\ket{\eta_{x1}}$ where $|\nu_{x0}|^2 = \frac{1}{2}(1-|\alpha_x|)$.
    
    \paragraph*{\bf Stage-2:} Now consider the registers $R_1R_{21}R_{22}R_3$ along with one of the $c\ln(\tfrac{1}{\delta^2\tau^2})$ independent copies. Neglect the superscript on the registers. 
    Notice that the simultaneous amplitude estimation is performed on the registers $R_1R_{21}R_{4}$. This operation can be given as $\sum_x\ketbra{x}\otimes AmpEst_x$ where $AmpEst_x$ uses the Grover iterator $G_x = -A_xU_{\overline{0}}A_xU_0$ and $A_x$ is the algorithm that acts as $A_x\ket{0}=\ket{x}$.
    Then, we obtain a state of the form 
    $$\ket{\psi_2} = \sum_{x\in [m]}\alpha_x\ket{x,\xi_x}\ket{\nu_x}\ket{\tau_1}\Big(\beta_{x,s}\ket{a_x}+\beta_{x,\overline{s}}\ket{E_x}\Big)\ket{0}$$ where 
    $\ket{a_x}$ is a normalized state of the form $\ket{a_x} = \gamma_{+}\ket{a_{x,+}} + \gamma_{-}\ket{a_{x,-}}$ that on measurement outputs $a \in \{a_{x,+}, a_{x,+}\}$ which is
    an $l$-bit string that behaves as $\displaystyle\left| \sin^2\left(\frac{a\pi}{2^l}\right) - |\nu_{x0}|^2 \right| \le \tfrac{1}{2^q}$.
    We denote the set $\{a_{x,+}, a_{x,-}\}$ by $\pmset_{a_x}$.
    
    \paragraph*{\bf Stages-3 to 5:} Let by $p_x$ and $p_{\tau}$ we denote $\nu_{x0}^2$ and $\frac{1}{2}(1-\tau) = \tring{\tau}$. Then from Proposition~\ref{prop:fhat-tau-scale-change}, we can observe that the only possible cases for any $x\in\{0,1\}^n$ are $p_x \le \tring{\tau}-\epsilon$ and $p_x > \tring{\tau}$.
    The proof of correctness for these cases follow directly from the proof for \etalgo.
    Hence, at the end of Stage-4 we have that for any $x$ such that $p_x> \tring{\tau}$, $$Pr\Big[R_{f}=\ket{1}\Big|R_1=\ket{x}\Big] \ge 1-\delta^2\tau^2 \ge 1-\delta,$$ and for any $x$ such that $p_x\le \tring{\tau}-\epsilon$ we have, $$Pr\Big[R_{f}=\ket{1}\Big|R_1=\ket{x}\Big] \le \delta^2 \tau^2 < \delta .$$
    So, using Proposition~\ref{prop:fhat-tau-scale-change}, we get that for any $x$ such that $|\alpha_x| < \tau$, $$Pr\Big[R_{f}=\ket{0}\Big|R_1=\ket{x}\Big] < \delta,$$ and for any $x$ such that $|\alpha_x| \ge \tau+2\epsilon$ we have, $$Pr\Big[R_{f}=\ket{0}\Big|R_1=\ket{x}\Big] \ge 1-\delta.$$
    
    Therefore, the overall probability of obtaining $\ket{1}$ in $R_{f}$ after stage-5 can be expressed as
    $$Pr\Big[R_{f}=\ket{0}\Big] \ge \sum_{x\in G}|\alpha_x|^2\cdot (1-\delta) \ge (\tau+2\epsilon)(1-\delta) \ge \tau/2$$ assuming that $\delta < \tfrac{1}{2}$.
    
    For the query complexity of this algorithm, we use $O(\frac{1}{\epsilon})$ queries in $SAE$ and perform the estimation in a total of $O(\log(\frac{1}{\delta\tau}))$ independent copies.
    In the last stage, the number of iterations of amplitude amplification done is $O(\frac{1}{\tau})$.
    Hence, we have to query complexity as $O(\frac{1}{\epsilon\tau}\log(\frac{1}{\delta\tau}))$ queries.
    
\end{proof}

\section{Choice of oracles}
\label{subsec:oracle}

Bravyi et al.~\cite{Bravyi2011QuantumDistributions} worked on designing quantum algorithms to analyse probability distributions induced by multisets. They considered an oracle, say $O_S$, to query an $n$-sized multiset, say $S$, in which an element can take one of $m$ values. Hence, the probabilities in the distribution of elements in those multisets are always multiples of $1/n$. They further proved that the query complexity of an algorithm in this oracle model is same as the sample complexity when sampled from the said distribution in a classical scenario. Li and Wu~\cite{Li2019QuantumEstimation} too used the same type of oracles for estimating entropies of a multiset.

We consider a general oracle in which the probabilities can be any real number, and are encoded in the amplitudes of the superposition generated by an oracle. We show below how to implement an oracle of our type for any distribution $D$, denoted $O_D$, using $O_S$.
\vspace*{-.5em}
\begin{equation}\label{eqn:OS_to_OD}
\ket{0^{\log(n)}} \ket{0^{\log(m)}} \xrightarrow{H^{\log(n)}} \tfrac{1}{\sqrt{n}} \sum_{i \in [n]} \ket{i} \ket{0^{\log(m)}} \xrightarrow{O_S} \tfrac{1}{\sqrt{n}} \sum_{i \in [n]} \ket{i} \ket{S_i} = \sum_{j \in [m]} \alpha_j \ket{\xi_j} \ket{j}
\end{equation}
It should be noted that one call to $O_D$ invokes $O_S$ only once.
Here the $\ket{\xi_j}$ states are normalized, and the probability of observing the second register is $|\alpha_j|^2$. Hence, ignoring the first register gives us the desired output of $O_D\ket{0^{\log(m)}}$ in the second register.

We use $O_D$ for \prob and \pmaxprob, and $O_S$ for the other array-based problems, namely, \Finf and variants of element distinctness.

\section{Algorithm for \pmaxprob problem}
\label{appendix:pmax}


\subsection{\pmaxprob problem with additive accuracy}

An algorithm for \promiseprob, with additive accuracy set to some $e$, is used to decide whether to search in the right half or the left half. 
%
It suffices to choose $k$ and $e$ such that $1/2^k \le \epsilon/2$ and $e \le \epsilon/2$ and repeatedly call the \promiseprob algorithm with accuracy $e$. Suppose $\tfrac{t}{2^k}$ is the threshold passed to the \promiseprob algorithm at some point. Then, if the algorithm returns {\tt TRUE} then $\pmax \ge \frac{t}{2^k}-e$, and so we continue to search towards the right of the current threshold; on the other hand if the algorithm returns {\tt FALSE} then $\pmax < \frac{t}{2^k}$, so we search towards its left. At the end some $t$ is obtained such that $\pmax \in [\tfrac{t}{2^k}-e,\tfrac{t+1}{2^k})$, an interval of length at most $\epsilon$. This is the idea behind the {\tt IntervalSearch} algorithm from our earlier work on non-linearity estimation~\cite[Algorithm~1]{Bera2021QuantumEstimation}. Once such a $t$ is obtained, $t/2^k$ can be output as an estimate of $\pmax$ which is at most $\epsilon$ away from the actual value. Lemma~\ref{lemma:pmax_lemma_additive} follows from Lemma~\ref{lemma:etalgo_correct} and the observation that $k$ binary searches have to be performed.

We now describe a quantum algorithm to estimate $\max_{x \in [m]} p_x=|\alpha_x|^2$ with an additive accuracy given a quantum black-box $O_D$ with the following behaviour.
$$O_D \ket{0^{\log(m)+a}} = \sum_{x \in \{0,1\}^{\log(m)}} \alpha_x \ket{x} \ket{\psi_x}$$
The black-box generates the distribution $\D=(p_x)_{x=1}^m$ when its first $\log(m)$ qubits are measured in the standard basis.

\pmaxadditive*

We design an algorithm namely {\tt IntervalSearch} to prove the lemma. The algorithm originally appeared in~\cite{Bera2021QuantumEstimation}. The idea behind Algorithm {\tt IntervalSearch} is quite simple. 
The algorithm essentially combines the \etalgo~with the classical binary search. 
Recall that given any threshold $\tau$, accuracy $\epsilon$ and error $\delta$, if \etalgo~outputs {\tt TRUE} then $\pmax\ge \tau-2\epsilon$ else if the output is {\tt FALSE} then $\pmax < \tau$.
The {\tt IntervalSearch} algorithm is as presented in Algorithm~\ref{algo:intervalsearch}.
\begin{algorithm}
\caption{Algorithm {\tt IntervalSearch} to find out an $\epsilon$-length
interval containing $\max_{i\in [n]}p_i$\label{algo:intervalsearch}}
\begin{algorithmic}
    \Require Distribution oracle $O_D$, size of the oracle $r=\log(m)+a$, size of the distribution $m$, accuracy $\epsilon$ and probability of error $\delta$
    \State Set $k = \left\lceil \log_2 \tfrac{1}{\epsilon}
    \right\rceil + 1$ \Comment{$k$ is the smallest integer {\it s.t.}
    $\tfrac{1}{2^k} \le \frac{\epsilon}{2}$; thus, $\frac{\epsilon}{4}
    < \frac{1}{2^k} \le \frac{\epsilon}{2}$}
    \State Set gap $g=\frac{\epsilon}{4}$
    \State Set boundaries $lower=\tfrac{1}{n}$, $upper=1$ and threshold
    $\tau=\frac{1}{2}$
    \For{$i=1 \ldots k$}
    \If{$\tau \le \epsilon$}
        \State Update $upper = \epsilon$
        \State \textbf{Break}
    \EndIf
	\If{\etalgo$(r, m, \tau, g, \frac{\delta}{k}) \to \mathtt{TRUE}$}
	    \State Update $lower = \tau - g$, $\tau = \tau +
    \tfrac{1}{2^{i+1}}$ ; $upper$ is unchanged
	\Else
	    \State Update $upper = \tau$, $\tau = \tau -
    \tfrac{1}{2^{i+1}}$; $lower$ is unchanged
	\EndIf
    \EndFor
    \State \Return $[lower,upper)$
\end{algorithmic}
\end{algorithm}

\begin{proof}
Notice that the interval $[lower, upper)$ at the start of the $i^{th}$ iteration is such that the size of the interval is either $\frac{1}{2^i}$ or $\frac{1}{2^i}-g$.
The algorithm essentially attempts to find a $\tau$ which is a multiple of $\frac{1}{2^k}$ in such a way that at $k-1^{th}$ iteration, $\tau$ is (almost) the center of an interval $J$ of size $\frac{1}{2^{k-1}}$ and $p_{max}\in J$.
It is clear that after the $k^{th}$ iteration the algorithm returns an interval of the form $[\frac{t}{2^k}-g, \frac{t+1}{2^k})$ for $t\in \{1, 2, \cdots 2^{k}-1\}$ and the length of the returned interval is at most $\frac{1}{2^k}+g \le \frac{\epsilon}{2}+g \le \epsilon$ as desired.
The correctness of the algorithm then follows from the correctness of \etalgo.
{\tt IntervalSearch} makes $k = O(\log(\frac{1}{\epsilon}))$ invocations of the \etalgo.

Since the accuracy parameter of each invocation of \etalgo~in {\tt IntervalSearch} is $\frac{\epsilon}{4}$ and the error parameter is $\frac{\delta}{k}$, from Lemma~\ref{lemma:etalgo_correct} we get that the query complexity of each invocation of \etalgo~is $O(\frac{1}{\epsilon\sqrt{\tau_i}}\log\log(\epsilon)\log(\frac{1}{\delta\tau_i}))$ where $\tau_i$ denotes the threshold at iteration $i$.
Hence, we get the total query complexity of {\tt IntervalSearch} as $\sum_{i=1}^k O(\frac{1}{\epsilon\sqrt{\tau_i}}\log(\frac{1}{\epsilon})\log(\frac{1}{\delta\tau_i}))$ which equals $O(\frac{1}{\epsilon\sqrt{\pmax}}\log{(\frac{1}{\epsilon})}\log(\frac{1}{\epsilon})\log{(\frac{1}{\delta\pmax})})$. The last equality uses the fact that $\tau_i \ge \pmax/2$ for any $i\in [k]$.
Now, since each time \etalgo is invoked with the error parameter $\frac{k}{\delta}$, using union bounds we can say that the {\tt IntervalSearch} algorithm returns an erred output with probability at most $\delta$.
\end{proof}

\begin{figure}
    \centering
    \includegraphics[width=0.6\linewidth]{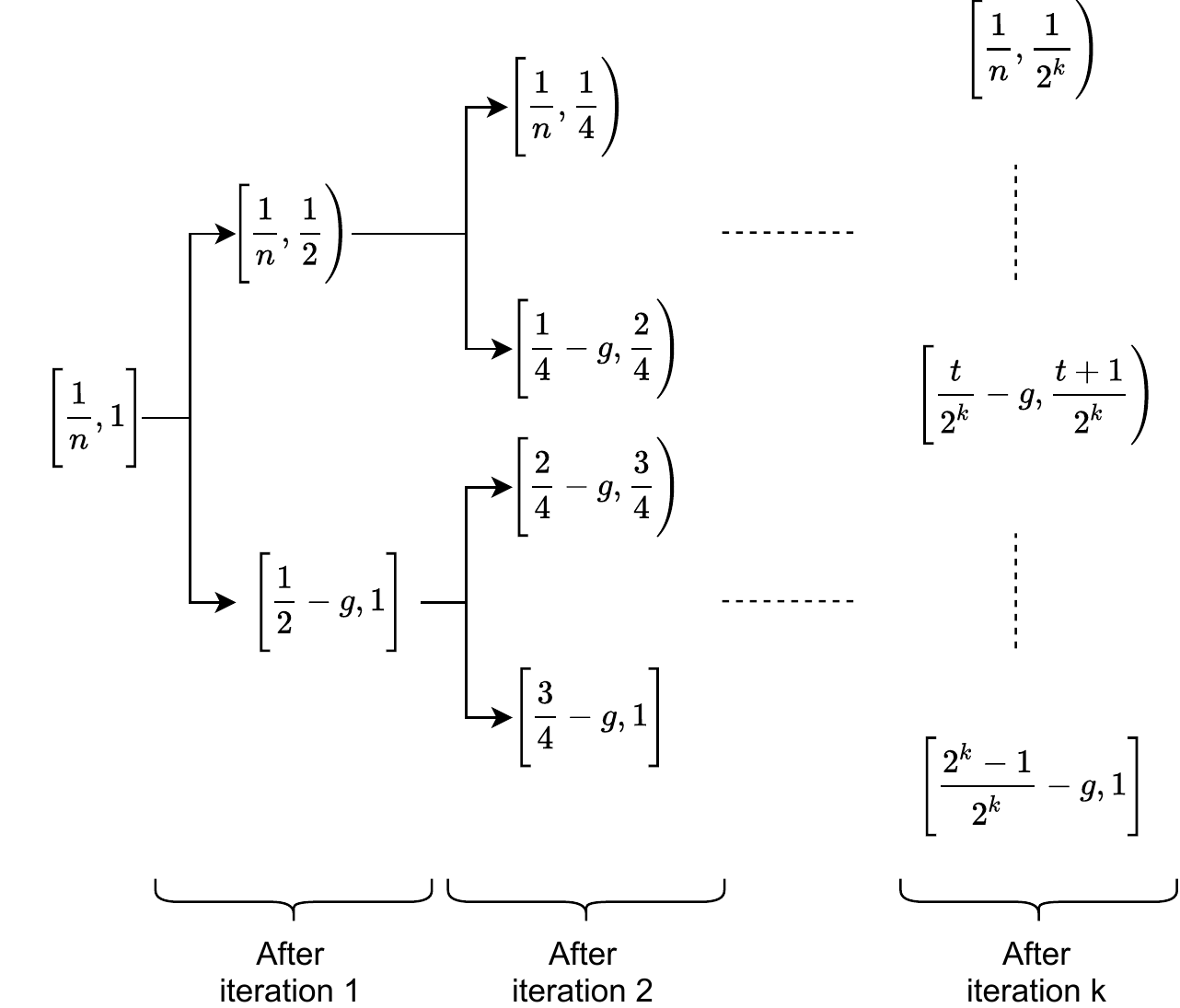}
    \caption{Illustration of the {\tt IntervalSearch} algorithm}
    \label{fig:my_label}
\end{figure}

\subsection{\pmaxprob problem with relative accuracy}\label{appendix:pmax_rel}

The algorithm for \pmaxprob with relative accuracy, denoted $\epsilon_r$, follows a similar idea as that of its additive accuracy version, except that it searches among the thresholds $1,(1-\epsilon_r),(1-\epsilon_r)^2, \ldots, (1-\epsilon_r)^{k-1}$ in which $k$ is chosen to be the smallest integer for which $(1-\epsilon_r)^{k-1} \le \tfrac{1}{m}$. Further, it calls the above algorithm for \promiseprob with relative error $\epsilon_r$. At the end of the binary search among the $k$ thresholds, we obtain some $t$ such that $\pmax \in [(1-\epsilon_r)(1-\epsilon_r)^{t+1},(1-\epsilon_r)^t)$. Clearly, if we output $(1-\epsilon_r)^t$ as the estimate $\widehat{\pmax}$, then $\pmax \le \widehat{\pmax}$ and $\pmax \ge (1-\epsilon_r)^2 \widehat{\pmax}$ as required.

Now, we present an algorithm to approximate $p_{max}$ with relative error.


\begin{lemmanonum}[Approximating $\pmax$ with relative error]
Given an oracle as required for the \prob problem, relative accuracy $\epsilon \in (0,1)$ and error $\delta$, there is a quantum algorithm that makes $\tilde{O}(\frac{m^{3/2}}{\epsilon})$ queries to the oracle and outputs an estimate $\tilde{p}_{max}$ such that with probability $1-\delta$, it holds that
$(1-\epsilon)\tilde{p}_{max} \le p_{max} < \tilde{p}_{max}$. The algorithm uses $\log(\frac{m}{\epsilon})+a)$ qubits.
\end{lemmanonum}

To solve the relative version of the \pmaxprob problem, we introduce a relative version of {\tt IntervalSearch} which we call {\tt IntervalSearchRel}. 
Similar to the {\tt IntervalSearch} algorithm, {\tt IntervalSearchRel} also combines the \etalgo with a classical binary search. But here the binary search is over the powers of $(1-\epsilon')$ where $\epsilon'= (1-\sqrt{1-\epsilon})$ rather than on intervals of length $\frac{1}{2^k}$.
The algorithm is as in Algorithm~\ref{algo:intervalsearchrel}.

\begin{algorithm}
\caption{Algorithm {\tt IntervalSearchRel} to return an $\epsilon$ relative estimate of $p_{max}$\label{algo:intervalsearchrel}}
\begin{algorithmic}
    \Require Distribution oracle $O_D$, size of the oracle $r=\log(m)+a$, size of the distribution $m$, accuracy $\epsilon$ and probability of error $\delta$
    \State Set $\epsilon' = 1-\sqrt{1-\epsilon}$ and set $k$ as the largest number {\it s.t.}
    $(1-\epsilon)^{2^k} \le \frac{1}{n} < (1-\epsilon')^{2^{k-1}}$
    \State Set boundaries $lower=\tfrac{1}{n}$, $upper=1$ and threshold
    $\tau=(1-\epsilon')^{\frac{2^k}{2}}$
    \For{$i=1 \ldots k$}
    \If{$\tau \le (1-\epsilon')$}
        \State Update $upper = (1-\epsilon')$
        \State \textbf{Break}
    \EndIf
	\If{\etalgo$(r, m, \tau, (\epsilon'\tau), \frac{\delta}{k}) \to \mathtt{TRUE}$}
	    \State Update $lower = (1-\epsilon')\tau$, $\tau = \tau/(1-\epsilon')^{\frac{2^k}{2^{i+1}}}$; $upper$ is unchanged
	\Else
	    \State Update $upper = \tau$, $\tau = \tau(1-\epsilon')^{\frac{2^k}{2^{i+1}}}$; $lower$ is unchanged
	\EndIf
    \EndFor
    \State \Return $p_{max} = upper$
\end{algorithmic}
\end{algorithm}

\begin{proof}
First observe that for any relative accuracy $\epsilon'$ and a threshold $\tau$, deciding the \prob problem with relative accuracy $\epsilon'$ is equivalent to deciding the additive \prob problem with additive accuracy $\epsilon'\tau$.
So, $\etalgo(r,m,\tau, (\epsilon'\tau),\frac{\delta}{k})$ even though is in additive terms, essentially solves the \prob problem with relative accuracy $\epsilon'$.
Next, in {\tt IntervalSearchRel}, at the end of $i^{th}$ iteration, any interval $[lower, upper)$ not on the extremes is of the form $\Big[(1-\epsilon')^{(\frac{(t+1)2^k}{2^{i}}+1)}, (1-\epsilon')^{(\frac{t2^k}{2^{i}})}\Big)$ where $t\in \{1,2,\cdots, 2^{i}-2\}$.
The left and the right extreme intervals are of the form $\Big[\frac{1}{m},(1-\epsilon')^{(1-\frac{1}{2^i})2^k}\Big)$ and $\Big[(1-\epsilon')^{(\frac{2^k}{2^{i}}+1)}, 1\Big]$ respectively.
In contrast to {\tt IntervalSearch}, by the end of $k-1^{th}$ iteration {\tt IntervalSearchRel} tries to find a $\tau$ which is a power of $(1-\epsilon')^{\frac{2^k}{2^{k-1}}} = (1-\epsilon')^{2}$ such that $\tau$ lies strictly inside an interval $J = \Big[(1-\epsilon')^{(2t+3)}, (1-\epsilon')^{2t}\Big)$ where $J$ contains $p_{max}$.
At the end of $k^{th}$ iteration, the interval $I = [lower, upper)$ is of the form $\Big[(1-\epsilon')^{(t+2)}, (1-\epsilon')^{t}\Big)$ where $t\in \{1,2,\cdots, 2^{k}-2\}$ and $p_{max}\in I$ if $I$ is not in the extremes.
The left and the right extreme intervals are of the form $\Big[\frac{1}{m}, (1-\epsilon')^{2^k-1}\Big)$ and $\Big[(1-\epsilon')^2, 1\Big]$ respectively.
The algorithm then returns $\tilde{p}_{max} = upper$. Now, since we know that $p_{max}\in I$, we have that $p_{max} < upper = \tilde{p}_{max}$ and $p_{max} \ge lower \ge (1-\epsilon')^2\tilde{p}_{max} = (1-\epsilon)\tilde{p}_{max}$.
So, we have $(1-\epsilon)\tilde{p}_{max} \le p_{max} < \tilde{p}_{max}$ as required.

At iteration $i$, as the accuracy parameter and the error parameter or \etalgo in {\tt IntervalSearchRel} are $\epsilon'\tau_i$ and $\frac{\delta}{k}$ , \etalgo makes $O(\frac{1}{\epsilon(\tau_i)^{3/2}}\log(\frac{1}{\epsilon\tau_i})\log(\frac{k}{\delta\tau_i}))$ queries to the oracle.
So, algorithm {\tt IntervalSearchRel} makes $\sum_{i=1}^k O(\frac{1}{\epsilon(\tau_i)^{3/2}}\log(\frac{1}{\epsilon\tau_i})\log(\frac{k}{\delta\tau_i}))$ \\$= O\Big(\frac{m^{3/2}}{\epsilon}\log(\frac{m}{\epsilon})\Big(\log(\log\log(\frac{1}{m})-\log\log(1-\epsilon)) + \log(\frac{1}{\delta\tau_i})\Big)\Big) = \tilde{O}(\frac{m^{3/2}}{\epsilon})$ queries to the oracle.
The error analysis simply follows from the union bound of errors at each iteration.
\end{proof}



\section{Simultaneous Amplitude Estimation and Hadamard Test}
\label{appendix:sae}

\subsection{Simultaneous Amplitude Estimation}

Let $[N] = \{y : 0\le y < 2^n-1 = N-1\}$ be an index set for some $n\in \mathbb{N}$. Let $A$ be an algorithm defined as $A = \sum_y \ketbra{y} \otimes A_y$ where $A_y$s are algorithms indexed by $y$ and all of which use some oracle $O$. Also let the number of times $O$ is called in any $A_y$ is $k$.
Then for each $y$, $A_y$ can be given as $A_y = U_{(k,y)} O U_{(k-1,y)} \cdots U_{(1,y)} O U_{(0,y)}$ with suitable $U_{(i,y)}$ unitaries.
Let the action of $A_y$ on $\ket{0}$ be defined as $A_y\ket{0} = \beta_{0y} \ket{0} + \beta_{1y}\ket{1}$ (This can also be easily generalized if $A_y$s are $n$ qubit algorithms.)
So, the action of $A$ on a state of the form $\sum_y\alpha_y\ket{y}\ket{0}$ can be given as
$$A\sum_y\alpha_y\ket{y}\ket{0} = \sum_y \alpha_y \ket{y} \big(\beta_{0y}\ket{0} + \beta_{1y}\ket{1}\big) = \sum_y \alpha_y \ket{y} \ket{\xi_y}(\text{say}) = \ket{\Psi}.$$
Now, without loss of generality assume that $\ket{0}$ is the good state and our objective is to obtain the estimates of $\beta_{0y}$s in parallel using some extra ancilla qubits, i.e, we would like to obtain a state of the form 
$$\ket{\Phi} = \sum_y \alpha_y\ket{y}\ket{\xi_y}\ket{\tilde{\beta}_{0y}}$$ where, for each $y$, $\sin^2\Big(\frac{\tilde{\beta}_{0y} \pi}{2^m}\Big) = \Breve{\beta}_{0y}$ is an estimate of $\beta_{0y}$ such that $|\Breve{\beta_{0y}}-\beta_{0y}|\le \epsilon$ for some given $0<\epsilon\le 1$.
We call this problem of simultaneous estimation of $\beta_{0y}s$ as \simulaeprob problem.

\begin{restatable}[\simulaeprob]{problem}{simulae_prob}
\label{lemma:simulae_prob}
Given an indexed set of algorithm $\{A_y:y\in[N]\}$ that can be described as $A_y = U_{(k,y)} O U_{(k-1,y)} \cdots U_{(1,y)} O U_{(0,y)}$ for some fixed $k$ and act as $A_y\ket{0} = \beta_{0y}\ket{0} + \beta_{1y}\ket{1},$ along with parameter $\epsilon$ and an algorithm $A_{initial}$ that produces the initial state $\ket{\Psi}=\sum_x\alpha_y\ket{y}\ket{0^m}$, output the state 
$\ket{\Phi} = \sum_i\alpha_y \ket{y}\ket{\tilde{\beta}_{0,y}}$ such that $|\beta_{0y}-\sin^2\big(\frac{\tilde{\beta}_{0y}\pi}{2^m}\big)|\le \epsilon$.
\end{restatable}

A naive approach to solve this problem would be to perform amplitude estimation of the state $\ket{\xi_y}$ conditioned on the first register being in $\ket{y}$ for each individual $y$.
Then, the total number of queries to the oracle $O$ would be $O(\frac{Nk}{\epsilon})$ where $O(k/\epsilon)$ is the query complexity due to a single amplitude estimation. However, this is very costly.
We give an algorithm that performs the same task but with just $O(\frac{k}{\epsilon})$ queries to the oracle $O$.

We denote the amplitude estimation algorithm due to Brassard et al.~\cite{brassard2002quantum} as $AmpEst$. The amplitude estimation algorithm to obtain an estimate with $m$ bits of precision can be given as $AmpEst = (F_m^{-1} \otimes \iden)\cdot \Lambda_m(G) \cdot (F_m \otimes \iden)$ where $F_m$ is the Fourier transform on $m$ qubits, $\Lambda_m(G)$ is the conditional operator defined as $\sum_x \ketbra{x}\otimes G^x$, $G = -AS_0AS_{\chi}$ is the Grover operator and $G^x$ implies that the $G$ operator is applied $x$ times in succession.
Also let $AmpEst_y$ be defined as $AmpEst = (F_m^{-1} \otimes \iden)\cdot \Lambda_m(G_y) \cdot (F_m \otimes \iden)$ where $G_y = -A_yS_0A_y^{\dagger}S_\chi$. Then notice that $\ket{\Phi}$ can be obtained from $\ket{\Psi}$, as $$\ket{\Phi} = \Big(\sum_y \ketbra{y}\otimes AmpEst_y\Big)\cdot \ket{\Psi}\ket{0^m}.$$
By $\mathbf{U}$ we denote the operator $\sum_y \ketbra{y}\otimes AmpEst_y$. We show that $\mathbf{U}$ can be implemented using $O(k\cdot 2^m) = O(k/\epsilon)$ queries to the oracle $O$. 

\begin{algorithm}[!h]
	\caption{\label{algo:simul_ae_algo}Simultaneous Amplitude Estimation Algorithm}
	\begin{algorithmic}[1]
	    \Require Oracle $O$, the set of indexed algorithms $\{A_y\}$, the algorithm $A_{initial}$, accuracy $\epsilon$ and error $\delta$.
	    \State Set $m = \lceil\frac{1}{\epsilon}\rceil+3$.
	    \State Initialize the three register state $R_1R_2R_3 = \ket{0^n}\ket{0}\ket{0^m}$.
	    \State Apply $A_{initial}$ on $R_1$.
	    \State Apply the quantum Fourier transform (QFT) $F_m$ on $R_3$.
	    \For{$i$ in $1$ to $m$, conditioned on $i^{th}$ qubit of $R_3$ being in $\ket{1}$, for $2^i$ many times}
        	    \For{$j$ in $1$ to $k-1$}
            	    \For{$y$ in $0$ to $N-1$}
            	        \State Apply $U_{(j,y)}$ on $R_2$ conditioned on $R_1$ being $\ket{y}$.
            	    \EndFor
            	    \State Apply $O$ on $R_2$.
            	\EndFor
            	\For{$y$ in $0$ to $N-1$}
            	        \State Apply $U_{(k,y)}$ on $R_2$ conditioned $R_1$ being $\ket{y}$.
        	        \EndFor
        \EndFor
        \State Apply the inverse QFT $F_m^{-1}$ on $R_3$.
        \State \Return $R_1R_2R_3$.
	\end{algorithmic}
    \end{algorithm}


\begin{restatable}[Simultaneous Amplitude Estimation]{theorem}{simulae_thm}
\label{theorem:simulaetheorem}
    Given an oracle $O$, a description of an algorithm $A = \sum_y\ket{y}\bra{y}\otimes A_y$ as defined earlier, an accuracy parameter $\epsilon$ and an error parameter $\delta$, \simulalgo uses $O(\frac{k}{\epsilon})$ queries to the oracle $O$ and with probability at least $1-\delta$ outputs a state of the form:
    $$\ket{\Phi} = \sum_y \alpha_y\ket{y}\ket{\xi_y}\ket{\tilde{\beta}_{0y}}$$ where $\sin^2\big(\frac{\tilde{\beta}_{0y}\pi}{2^m}\big)=\Breve{\beta_{0y}}$ is an $\epsilon$-estimate of $\beta_{0y}$ for each $y$.
\end{restatable}

Before we proceed to prove Theorem~\ref{theorem:simulaetheorem}, consider the following lemmas which would be useful in proving Theorem~\ref{theorem:simulaetheorem}. Here $\mbc{i,p}(U)$ denotes the operator $\iden^{i-1}\otimes \ketbra{p}\otimes \iden^{m-i} \otimes U$.

\begin{lemma}\label{lemma:product_distributive}
    Let $\{A_y\}$ and $\{B_y\}$ be two sets of indexed unitaries. Then, $$\sum_y\ketbra{y} \otimes (A_y\circ B_y) = \Big(\sum_y\ketbra{y} \otimes (A_y)\Big)\circ \Big(\sum_z\ketbra{z} \otimes (B_z)\Big).$$
\end{lemma}
\begin{proof}
    \begin{align*}
        \Big(\sum_y\ketbra{y} \otimes A_y\Big)\circ \Big(\sum_z\ketbra{z} \otimes B_z\Big) &=\sum_{y,z(}\ketbra{y}\circ\ketbra{z}) \otimes (A_y\circ B_z)\\
        &=\sum_y\ketbra{y}\otimes (A_y\circ B_y)
    \end{align*}
\end{proof}

\begin{lemma}\label{lemma:cproduct_1}
    For any two unitaries $A$ and $B$, we have $$\mbc{i,p}(A)\circ \mbc{i,q}(B) = \delta_{p,q}\cdot \mbc{i,p}(A\circ B)$$ where $\delta_{p,q} = 1$ if $p=q$ and $0$ otherwise.
\end{lemma}
\begin{proof}
    \begin{align*}
        \mbc{i,p}(A)\circ \mbc{i,q}(B) &= \big(\iden^{i-1}\otimes \ketbra{p}\otimes \iden^{m-i} \otimes A\big)\circ \big(\iden^{i-1}\otimes \ketbra{q}\otimes \iden^{m-i} \otimes B\big)\\
        &= \iden^{i-1}\otimes (\ketbra{p}\circ\ketbra{q}) \otimes \iden^{m-i} \otimes (A\circ B)\\
        &= \iden^{i-1}\otimes \delta_{p,q}(\ketbra{p}) \otimes \iden^{m-i} \otimes (A\circ B)\\
        &= \delta_{p,q} \big(\iden^{i-1}\otimes \ketbra{p} \otimes \iden^{m-i} \otimes (A\circ B)\big)\\
        &= \delta_{p,q}\cdot\mbc{i,p}(A\circ B)
    \end{align*}
\end{proof}

\begin{lemma}
    For any two unitaries $A$ and $B$, we have    $\sum_{y}\ketbra{y}\otimes \big[\mbc{i,1}(A\circ B)+\mbc{i,0}(\iden)\big] = \Big\{\sum_{y}\ketbra{y}\otimes \big[\mbc{i,1}(A)+\mbc{i,0}(\iden)\big]\Big\}\circ \Big\{\sum_{y}\ketbra{y}\otimes \big[\mbc{i,1}(B)+\mbc{i,0}(\iden)\big]\Big\}$
\end{lemma}
\begin{proof}
    \begin{align*}
        &\sum_{y}\ketbra{y}\otimes \big[\mbc{i,1}(A\circ B)+\mbc{i,0}(\iden)\big]\\ &=\sum_{y}\ketbra{y}\otimes \big[\big(\mbc{i,1}(A)\circ \mbc{i,1}(B)\big)+\mbc{i,0}(\iden)\big]~\text{(Using Lemma~\ref{lemma:cproduct_1})}\\
        &= \sum_{y}\ketbra{y}\otimes \big(\big[\mbc{i,1}(A) + \mbc{i,0}(\iden)\big]\circ \big[\mbc{i,1}(B)+\mbc{i,0}(\iden)\big]\big)~\text{(Using Lemma~\ref{lemma:cproduct_1})}\\
        &= \Big\{\sum_{y}\ketbra{y}\otimes \big[\mbc{i,1}(A) + \mbc{i,0}(\iden)\big]\Big\}\circ \Big\{\sum_{y}\ketbra{y}\otimes \big[\mbc{i,1}(B) + \mbc{i,0}(\iden)\big]\Big\}~\text{(Using Lemma~\ref{lemma:product_distributive})}
    \end{align*}
\end{proof}

\begin{proof}[Proof of Theorem~\ref{theorem:simulaetheorem}]
First, on applying $A_{initial}$ on $R_1$, we obtain the state, $\ket{\Phi} = \sum_y\alpha_y\ket{y}$.
Now, let $\mathbf{U}$ denote the operator $\sum_y \ketbra{y}\otimes AmpEst_y$. 
Since $A_{initial}$ does not use oracle $O$, it suffices to show that $\mathbf{U}$ can be implemented with $O(k/\epsilon)$ queries to the oracle $O$.
Using Lemma~\ref{lemma:product_distributive} the operator $\mathbf{U}$ can be written as 
\begin{align*}
    \mathbf{U} &= \sum_y \ketbra{y}\otimes AmpEst_y\\
    &= \sum_y \ketbra{y}\otimes \Big((F_m^{-1} \otimes \iden)\cdot \Lambda_m(G_y) \cdot (F_m \otimes \iden)\Big)\\
    &= \Big(\sum_y \ketbra{y}\otimes (F_m^{-1} \otimes \iden)\Big)\circ \Big(\sum_y \ketbra{y}\otimes \Lambda_m(G_y)\Big)\circ \Big(\sum_y \ketbra{y}\otimes (F_m \otimes \iden)\Big)\\
    &= \Big(\iden^n\otimes (F_m^{-1} \otimes \iden)\Big)\circ \Big(\sum_y \ketbra{y}\otimes \Lambda_m(G_y)\Big)\circ \Big(\iden^n\otimes (F_m \otimes \iden)\Big)\\
    &= \Big(\iden^n\otimes (F_m^{-1} \otimes \iden)\Big)\circ \Big(\sum_y \ketbra{y}\otimes \sum_x \ketbra{x}\otimes G_y^x\Big)\circ \Big(\iden^n\otimes (F_m \otimes \iden)\Big)
\end{align*}
Notice that the middle operator in the above equation can be rephrased as:
\begin{align}
    &\sum_y \ketbra{y}\otimes \sum_x \ketbra{x}\otimes G_y^x\nonumber\\ 
    =& \prod_{i=1}^m\bigg[\sum_y \ketbra{y}\otimes \bigg(\Big[\iden^{i-1}\otimes \ketbra{1}\otimes \iden^{m-i} \otimes G_y^{2^i}\Big] + \Big[\iden^{i-1}\otimes \ketbra{0}\otimes \iden^{m-i} \otimes \iden \Big]\bigg)\bigg] \label{eqn:total_grover}
\end{align}

Now see that for any $i$,
\begin{align}
    &\sum_y \ketbra{y}\otimes \bigg(\Big[\iden^{i-1}\otimes \ketbra{1}\otimes \iden^{m-i} \otimes G_y^{2^i}\Big] + \Big[\iden^{i-1}\otimes \ketbra{0}\otimes \iden^{m-i} \otimes \iden \Big]\bigg)\nonumber\\
    =&\sum_y \ketbra{y}\otimes \Big(\mbc{i,1}\Big(G_y^{2^i}\Big) + \mbc{i,0}(\iden) \Big)\label{eqn:mid-operator-one-term}\\
    =&\sum_y \ketbra{y}\otimes \Big(\mbc{i,1}\Big((-A_yS_0A_y^{\dagger}S_{\chi})^{2^i}\Big) + \mbc{i,0}(\iden) \Big)\nonumber\\
    =& \Bigg[\bigg\{\sum_y \ketbra{y}\otimes \Big(\mbc{i,1}\big(-A_y\big) + \mbc{i,0}(\iden) \Big)\bigg\} \circ 
    \bigg\{\sum_y \ketbra{y} \otimes \Big(\mbc{i,1}\big(S_0\big) + \mbc{i,0}(\iden) \Big)\bigg\} ~\circ\nonumber\\
    &\hspace{1cm} \bigg\{\sum_y \ketbra{y}\otimes \Big(\mbc{i,1}\big(A_y^{\dagger}\big) + \mbc{i,0}(\iden) \Big)\bigg\} \circ 
    \bigg\{\sum_y \ketbra{y} \otimes \Big(\mbc{i,1}\big(S_{\chi}\big) + \mbc{i,0}(\iden) \Big)\bigg\}\Bigg]^{2^i}\nonumber\\
    =& \Bigg[\bigg\{\sum_y \ketbra{y}\otimes \Big(\mbc{i,1}\big(-A_y\big) + \mbc{i,0}(\iden) \Big)\bigg\} \circ 
    \bigg\{\iden^n \otimes \Big(\mbc{i,1}\big(S_0\big) + \mbc{i,0}(\iden) \Big)\bigg\} ~\circ\nonumber\\
    &\hspace{1cm} \bigg\{\sum_y \ketbra{y}\otimes \Big(\mbc{i,1}\big(A_y^{\dagger}\big) + \mbc{i,0}(\iden) \Big)\bigg\}\circ 
    \bigg\{\iden^n \otimes \Big(\mbc{i,1}\big(S_{\chi}\big) + \mbc{i,0}(\iden) \Big)\bigg\}\Bigg]^{2^i}\label{eqn:mid-operator-one-term-equal}
\end{align}


Next, since we have $A_y = U_{(k,y)}OU_{(k-1,y)}\cdots U_{(1,y)}OU_{(0,y)}$, we can write
\begin{align}
    & \sum_y \ketbra{y}\otimes \Big(\mbc{i,1}\big(-A_y\big) + \mbc{i,0}(\iden) \Big)\label{eqn:controlled-algo}\\
    =& -\sum_y \ketbra{y}\otimes \Big(\mbc{i,1}\big(U_{(k,y)}OU_{(k-1,y)}\cdots U_{(1,y)}OU_{(0,y)}\big) + \mbc{i,0}(\iden) \Big)\nonumber\\
    =& -\prod_{j=k}^{1}\Bigg[\bigg\{\sum_y \ketbra{y}\otimes \bigg( \mbc{i,1}\big(U_{(j,y)}\big) + \mbc{i,0}(\iden)\bigg)\bigg\}\circ\bigg\{\sum_y \ketbra{y}\otimes \bigg(\mbc{i,1}(O) + \mbc{i,0}(\iden)\bigg)\bigg\}\Bigg] \circ\nonumber\\
    & \hspace{10cm}\sum_y \ketbra{y}\otimes \bigg(\mbc{i,1}\big(U_{(0,y)}\big) + \mbc{i,0}(\iden)\bigg)\nonumber\\
    =& -\prod_{j=k}^{1}\Bigg[\bigg\{\sum_y \ketbra{y}\otimes \bigg( \mbc{i,1}\big(U_{(j,y)}\big) + \mbc{i,0}(\iden)\bigg)\bigg\}\circ\bigg\{\iden^{n}\otimes \bigg(\mbc{i,1}(O) + \mbc{i,0}(\iden)\bigg)\bigg\}\Bigg] \circ\nonumber\\
    & \hspace{10cm}\sum_y \ketbra{y}\otimes \bigg(\mbc{i,1}\big(U_{(0,y)}\big) + \mbc{i,0}(\iden)\bigg)\nonumber\label{eqn:final_sequence}\\
\end{align}
Each of the $\Big[\sum_y \ketbra{y}\otimes \Big( \mbc{i,1}\big(U_{(j,y)}\big) + \mbc{i,0}(\iden)\Big)\Big]$ terms can be implemented as 
$$\prod_{y=0}^{N-1} \bigg\{\Big[\ketbra{y}\otimes \Big( \mbc{i,1}\big(U_{(j,y)}\big) + \mbc{i,0}(\iden)\Big)\Big] + \sum_{x\neq y}\ketbra{x}\otimes \iden^m \otimes \iden\bigg\}$$ which can be identified as a sequence of $N$ controlled gates that do not use any queries to the oracle $O$.
Next notice that for each $i$, the operator $\Big[\iden^{n}\otimes \Big(\mbc{i,1}(O) + \mbc{i,0}(\iden)\Big)\Big]$ is applied independent of the state in the first register.
So, this operator can be implemented as a single controlled-oracle operation that uses 1 oracle query. With that we can see that the number of oracle queries required to implement $\Big[\sum_y \ketbra{y}\otimes \Big(\mbc{i,1}\big(-A_y\big) + \mbc{i,0}(\iden) \Big)\Big]$ (operator in Equation~\ref{eqn:controlled-algo}). is exactly $k$.

Using similar analysis for the operator $\Big[\sum_y \ketbra{y}\otimes \Big(\mbc{i,1}\big(A_y^{\dagger}\big) + \mbc{i,0}(\iden) \Big)\Big]$, we can see that the required number of oracle queries required to implement this operator is $k$.
Now, using the equivalence between the operators in Equation~\ref{eqn:mid-operator-one-term} and Equation~\ref{eqn:mid-operator-one-term-equal}, the total number of oracle queries required for the operation in Equation~\ref{eqn:mid-operator-one-term}, can be calculated as $2k\cdot {2^i}$ since the controlled-Grover operator is applied $2^i$ times.
This in turn implies that the total number of calls to oracle $O$ that is required to implement the operation in Equation~\ref{eqn:total_grover} is $\sum_{i=1}^{m} 2k\cdot{2^i} = O(k\cdot2^m)$. Since, we have set $m=O(1/\epsilon)$ we get the query complexity of $\simulalgo$ as $O(k/\epsilon)$.
\end{proof}

\subsection{Hadamard test for inner product estimation}

Suppose that we have two algorithms $A_{\psi}$ and $A_{\phi}$ that generate the state $A_{\psi}\ket{0^n}=\ket{\psi}$ and $A_{\phi}\ket{0^n}=\ket{\phi}$ respectively. 
Our task is to return an estimate to $|\bra{\psi}\ket{\phi}|$ with $\epsilon$ accuracy.
Since, we have description of both $A_{\psi}$ and $A_{\phi}$, it is quite straightforward to estimate the probability of obtaining $\ket{0^n}$ in the state $A_{\psi}^{\dagger}A_{\phi}\ket{0^n}$ with $\epsilon^2$ accuracy from which one can obtain an estimate of $|\bra{\psi}\ket{\phi}|$ with $\epsilon$ accuracy.
The query complexity of such an algorithm would be $O(1/\epsilon^2)$.
We show that obtaining such an estimate is possible with just $O(1/\epsilon)$ queries to $A_{\phi}$ and $A_{\psi}$.

Now, consider the following algorithm:
    \begin{algorithm}[!h]
	\caption{\label{algo:inner_prod_est}Algorithm for estimating inner product}
	\begin{algorithmic}[1]
	    \Require Controlled versions of $A_{\psi}$ and $A_{\phi}$.
	    \State Initialize the two register state $R_1R_2 = \ket{0}\ket{0^n}$.
	    \State Apply $H$ on $R_1$.
	    \State Controlled on $R_1$ being in $\ket{0}$, apply $A_{\psi}$ on $R_2$.
	    \State Controlled on $R_1$ being in $\ket{1}$, apply $A_{\phi}$ on $R_2$.
	    \State Apply $H$ on $R_1$.
	    \State Estimate the probability of obtaining $\ket{0}$ on measuring $R_1$ with $\epsilon/2$ accuracy. Let the estimate be $\tau$.
	    \State \Return $2\tau-1$.
	\end{algorithmic}
    \end{algorithm}

\begin{proof}[Proof of Algorithm~\ref{algo:inner_prod_est}]
    The state evolution in Algorithm~\ref{algo:inner_prod_est} can be seen as follows:
\begin{align*}
    \ket{0}\ket{0^n} & \xrightarrow{H\otimes I} \frac{1}{\sqrt{2}}\big(\ket{0}\ket{0^n} + \ket{1}\ket{0^n}\big)\\
    & \xrightarrow{\ket{0}\bra{0}\otimes A_{\psi} + \ket{1}\bra{1} \otimes I} \frac{1}{\sqrt{2}}\big(\ket{0}\ket{\psi} + \ket{1}\ket{0^n}\big)\\
    & \xrightarrow{\ket{0}\bra{0}\otimes I + \ket{1}\bra{1} \otimes A_{\phi}} \frac{1}{\sqrt{2}}\big(\ket{0}\ket{\psi} + \ket{1}\ket{\phi}\big)\\
    & \xrightarrow{H\otimes I} \frac{1}{2} \Big[\ket{0}\big(\ket{\psi}+\ket{\phi}\big) + \ket{1}\big(\ket{\psi}-\ket{\phi}\big)\Big]\\
\end{align*}

    The probability of measuring the $R_1$ register as $\ket{0}$ in the final state can be calculated as
    $$Pr\big[\ket{0}_{R_1}\big] = \Big\| \frac{1}{2} \big(\ket{\psi} + \ket{\phi}\big) \Big\|^2 = \frac{1}{2}\big(1-\big| \bra{\psi}\ket{\phi} \big|\big).$$
    Observe that to obtain $|\bra{\psi}\ket{\phi}|$ with $\epsilon$ accuracy, it suffices to estimate $\frac{1}{2}\big(1-|\bra{\psi}\ket{\phi}|\big)$ with $\epsilon/2$ accuracy which can be performed by the quantum amplitude amplification algorithm using $O(1/\epsilon)$ queries to $A_{\phi}$ and $A_{\psi}$.
\end{proof}

\section{Non-linearity Estimation}
\label{sec:non-lin-est}

The non-linearity estimation problem is essentially the amplitude version of the \pmaxprob problem with the Deutsch-Jozsa circuit as the oracle $O_D$.
Combining the algorithm for \amprob with the intervalsearch algorithm, we obtain the following lemma.

\begin{lemma}
    \label{lemma:non-lin}
    Given a Boolean function $f:\{0,1\}^n\xrightarrow{}\{0,1\}$ as an oracle, an accuracy parameter $\lambda$ and an error parameter $\delta$, there exists an algorithm that returns an estimate $\tilde{\eta}_f$ such that $|\eta_f - \tilde{\eta}_f|\le \lambda$ with probability at least $1-\delta$ using $O(\frac{1}{\lambda\hat{f}_{max}}\log(\frac{1}{\lambda})\log(\frac{1}{\delta\hat{f}_{max}}))$ queries to the oracle of $f$. 
\end{lemma}

The proof of Lemma~\ref{lemma:non-lin} follows similar to the proof of Algorithm~\ref{algo:intervalsearch} for $\pmax$.
The total query complexity can be obtained as $\sum_{i=1}^k O\Big(\frac{1}{\lambda\tau_i}\log(\frac{1}{\lambda})\log(\frac{1}{\delta\tau_i})\Big) = O\Big(\frac{1}{\lambda\hat{f}_{max}}\log(\frac{1}{\lambda})\frac{1}{\delta\hat{f}_{max}}\Big)$ queries using query complexity of \highampalgo and the fact that $\tau_i \ge \hat{f}_{max}/2$ for any $i$.

\section{Application of \prob for $k$-Distinctness}
\label{sec:gkd}




In~\cite{Montanaro2016TheMoments}, Montanaro hinted at a possible algorithm for the promise problem \dgkd by reducing it to the \Finf problem~\footnote{His reduction was to a relative-gap version of \gkd; however, the same idea works for the additive-gap version that we consider in this paper.}. The idea is to estimate the modal frequency of an array $A$ up to an additive accuracy $\Delta/2$ and then use this estimate to decide if there is some element of $A$ with frequency at least $k$. The query complexity would be same as that of \Finf.

Here we show a reduction from \dgkd to a promise version of \prob which allows us to shave off a $\log(\tfrac{n}{\Delta})$ factor from the above complexity. For \dgkd we are given an oracle $O_S$ to access the elements of $A$. First use $O_S$ to implement an oracle $O_D$ for the distribution $\D=(p_i)_{i=1}^m$ induced by the frequencies of the values in $A$. Then call the algorithm for \promiseprob with threshold $k/n$ and additive accuracy $\Delta/n$. Now observe that if there exists some $i \in [1, \ldots m]$ whose frequency is at least $k$, then $p_i \ge \tfrac{k}{n}$, and the \promiseprob algorithm will return {\tt TRUE}. On the other hand, if the frequency of every element is less than $k-\Delta$, then for all $i$, $p_i < \tfrac{k}{n} - \tfrac{\Delta}{n}$; the \promiseprob algorithm will return {\tt FALSE}. The query complexity of this algorithm is $\Tilde{O}\left(\tfrac{1}{\Delta/n}\tfrac{1}{\sqrt{k/n}}\right)$ which proves Lemma~\ref{lemma:dgkd-ub}.
The space complexity is the same as that of solving \promiseprob problem.


As for Lemma~\ref{lemma:kd-ub}, it is easy to see that \kd is equivalent to \dgkd with $\Delta=1$ and so the above algorithm can be used.

\section{Application of \pmaxprob for \Finf}\label{sec:min_entropy_est}


To compute the modal frequency of an array $A$, given an oracle $O_S$ to it, we first use $O_S$ to implement $O_D$ whose amplitudes contain the distribution $D_A$ induced by the values of $A$: $D_A = (p_i)_{i=1}^m$ where $p_i = |\{i\in [n] : A[i]=x\}|/n$. Then we can use the algorithms for \pmaxprob for $O_D$. The estimate obtained from that algorithm has to rescaled by multiplying it by $n$ to obtain an estimate of the largest frequency of $A$. If we call the additive accuracy algorithm for \pmaxprob with accuracy set to $\epsilon/n$, then we get an estimate of \Finf with additive error $\epsilon$. No such scaling of the error is required if we call the relative accuracy algorithm for \pmaxprob to obtain an estimate of \Finf with relative error. Thus Lemma~\ref{lemma:finf-ub} is proved.

\section{Complexity analysis of \pmaxprob estimation by Li et al.~\cite{Li2019QuantumEstimation}}
\label{sec:li_wu_compare}

It is well known that the current best known algorithm for solving $k$-distinctness problem for any general $k$ is the quantum walk based algorithm due to Ambainis\cite{Ambainis2007QuantumDistinctness} which has a query complexity of $O(n^{k/k+1})$. 
Here, we show that using that quantum walk based algorithm, the query complexity of \pmaxprob estimation algorithm proposed in~\cite[Algorithm~7]{Li2019QuantumEstimation}, which we call \liwualgo, with $\epsilon$ relative error is in fact $O(n)$.
Theorem~7.1 of~\cite{Li2019QuantumEstimation} states that the quantum query complexity of approximating $\max_{i\in [n]}p_i$ within a multiplicative error $0<\epsilon\le 1$ with success probability at least $\Omega(1)$ using \liwualgo is the query complexity of $\frac{16\log(n)}{\epsilon^2}$-distinctness problem.

So we have the complexity of $\frac{16\log(n)}{\epsilon^2}$-distinctness as $n^{\frac{16\log(n)}{16\log(n)+\epsilon^2}}$.
Now,
$$\frac{16\log(n)}{16\log(n)+\epsilon^2} = \frac{\log(n)}{\log(n)+(\epsilon^2/16)}$$
Since we have $0 < \epsilon \le 1$, $\frac{\epsilon^2}{16}\le \frac{1}{16}$.
This gives us that $\frac{\log(n)}{\log(n)+(\epsilon^2/16)}\ge \frac{
\log(n)}{\log(n)+(1/16)} = 1- \frac{1}{16\log(n)+1}$.
So, we have $$n^{\frac{16\log(n)}{16\log(n)+\epsilon^2}} \ge n^{1-\frac{1}{16\log(n)+1}} \ge n^{1-\frac{1}{16\log(n)}} = \frac{n}{n^{\frac{1}{16\log(n)}}} = \frac{n}{e^{\frac{1}{16}}} \ge n/2.$$
The second last equality is due to the fact that $n^{\frac{1}{\log(n)}} = e$.
So for any relative error $\epsilon$, the algorithm makes $O(n)$ queries to the oracle.


\section{Reductions between problems}
\label{appendix:reductions}
In this section, we describe all the reductions between various problems encountered in this draft.

\noindent\textbf{\prob $\leqslant_T$ \pmaxprob:} Given \prob($O_D,\tau,\epsilon$), solve \pmaxprob($O_D,\epsilon/3$) and return {\tt TRUE} if $\tilde{p}_{max}\ge \tau-\frac{\epsilon}{2}$ else return {\tt FALSE}.
\vspace{10pt}

\noindent\textbf{\pmaxprob $\leqslant_T$ \prob:} Given \pmaxprob($O_D, \epsilon$) search for the largest integer $t\in \{1,2,\cdots,2^k\}$ such that \prob($O_D, \frac{t}{2^k}, \frac{\epsilon}{4}$) returns {\tt TRUE} where $k = \left\lceil \log(\frac{1}{\epsilon})+1 \right\rceil$ and return the interval $[\frac{t}{2^k}-\epsilon, \frac{t+1}{2^k})$ if $t\neq 2^k$ and return $[1-\frac{1}{2^k}, 1]$ if $t=2^k$.
The search is performed using binary search which imparts an additional $\log$ factor overhead to the complexity of solving \pmaxprob. See Section~\ref{appendix:pmax}
\vspace{10pt}

\noindent\textbf{\dgkd $\leqslant_T$ \prob:} Given \dgkd($O_A, \Delta$), solve \prob($O_A, \frac{k}{2}, \frac{\Delta}{3n}$) and return as \prob($O_A, \frac{\Delta}{n}$) returns. See Section~\ref{sec:gkd}.
\vspace{10pt}

\noindent\textbf{\dgkd $\leqslant_T$ \Finf:} Given \dgkd($O_A, \Delta$), solve \Finf($O_A, \frac{\Delta}{3n}$) and return {\tt TRUE} if $\tilde{f}_{inf} \ge k-\frac{\Delta}{2}$ else return {\tt FALSE}.
\vspace{10pt}

\noindent\textbf{\dgkd $\leqslant_T$ \kd:} Given \dgkd($O_A, \Delta$), solve \kd($O_A$) and return as \kd($O_A$) returns.
\vspace{10pt}

\noindent\textbf{\kd $\leqslant_T$ \dgkd:} Given \kd($O_A$), solve \dgkd($O_A, 1$) and return as \dgkd($O_A, 1$) returns.
\vspace{10pt}

\noindent\textbf{\Finf $\leqslant_T$ \pmaxprob:} Given \Finf($O_A,\epsilon$), solve \pmaxprob($O_A,\epsilon$) and return as \pmaxprob($O_A,\epsilon$) returns. See Section~\ref{sec:min_entropy_est}.
\vspace{10pt}

\noindent\textbf{\Finf $\leqslant_T$ \kd:} Given \Finf($O_A,\epsilon$), return the largest $k$ such that \kd($O_A$) returns {\tt TRUE}. The search is performed using a binary search which imparts an additional $\log$ factor overhead to the complexity of solving \Finf.
\vspace{10pt}

\noindent\textbf{\kd $\leqslant_T$ \Finf:} Given \kd($O_A$), solve \Finf($O_A, 1/3$) and return {\tt TRUE} if $\tilde{f}_{inf} \ge k-\frac{1}{2}$ else return {\tt FALSE}.
\vspace{10pt}

\noindent\textbf{\prob $\leqslant_T$ \amprob:} Given \prob($O_A, \tau,\epsilon$), solve \amprob($O_A, \sqrt{\tau}, \epsilon$) and return as \amprob($O_A, \sqrt{\tau}, \epsilon$) returns.
\vspace{10pt}






\end{document}